\documentclass[aps,prd,a4,amsmath,nofootinbib,preprint,superscriptaddress,dvipsnames]{revtex4-2}
\usepackage{graphicx}
\usepackage{color}
\usepackage{bbold}
\usepackage[makeroom]{cancel}
\usepackage{ulem}
\usepackage{epsfig}
\usepackage{epsf}
\usepackage{dcolumn}
\usepackage{bm}
\usepackage{soul}
\usepackage{hhline}
\usepackage{multirow}
\usepackage{dsfont}
\usepackage{braket}
\usepackage{slashed}
\usepackage{amssymb}
\usepackage{tikz,xcolor}
\usepackage[CJKbookmarks=true, colorlinks=true, linkcolor=blue, urlcolor=blue,citecolor=blue]{hyperref}
\usepackage[utf8]{inputenc}
\definecolor{mygreen}{rgb}{0, 0.7, 0.2}

\newcommand{\lam}{\Lambda}

\newcommand{\ulambda}{\underline{\lambda}}
\newcommand{\BB}{{\cal B}}
\newcommand{\HH}{{\cal H}}
\newcommand{\bem}{b^{\textrm{em}}}
\newcommand{\dd}{\text{d}}

\begin{document}

\title{Semileptonic decays of spin-entangled baryon--antibaryon pairs}
\author{Varvara Batozskaya}
\email[]{varvara.batozskaya@ncbj.gov.pl}
\affiliation{Institute of High Energy Physics,  Beijing 100049, People's Republic of China}
\affiliation{National Centre for Nuclear Research, Pasteura 7, 02-093 Warsaw, Poland}
\author{Andrzej Kupsc}
\affiliation{Department of Physics and Astronomy, Uppsala University, Box 516, SE-75120
  Uppsala, Sweden}
\affiliation{National Centre for Nuclear Research, Pasteura 7, 02-093 Warsaw, Poland}
\author{Nora Salone}
\affiliation{National Centre for Nuclear Research, Pasteura 7, 02-093 Warsaw, Poland}
\author{Jakub Wiechnik}
\affiliation{National Centre for Nuclear Research, Pasteura 7, 02-093 Warsaw, Poland}
\affiliation{Faculty of Physics, University of Warsaw, Pasteura 5, 02–093 Warsaw, Poland}
\begin{abstract}
A modular representation for the semileptonic decays of baryons originating from spin polarized and correlated baryon--antibaryon pairs is derived. The complete spin information of the decaying baryon is propagated to the daughter baryon via a real-valued matrix. It allows to obtain joint differential distributions in  sequential processes involving the semileptonic decay in a straightforward way.
The formalism is suitable for extraction of the semileptonic formfactors in experiments where  strange-baryon--antibaryon pairs are produced in electron--positron annihilation or in charmonia decays. We give examples such as the complete angular distributions in the $e^+e^-\to \lam\bar\lam$ process, where $\lam\to pe^-\bar{\nu}_e$ and $\bar\lam\to\bar{p}\pi^+$. The formalism can also be used to describe the distributions in semileptonic decays of charm and bottom baryons. Using the same principles, the modules to describe electromagnetic and neutral current weak baryon decay processes involving a charged lepton--antilepton pair can be obtained. As an example, we provide the decay matrix for the Dalitz transition between two spin-1/2 baryons.
\end{abstract}

\date{\today}
\maketitle

\section{Introduction}

Baryon semileptonic (SL) decays are an important tool to study transitions between ground state baryons. Comparing to the nonleptonic baryon decays where at least three hadronic currents are involved,
the SL transition involves only a two-point hadronic vertex and the external $W$-boson field coupled to the leptonic current. The properties of the hadronic vertices are described by a set of scalar functions, {\it formfactors}, that depend on the invariant mass squared of the emitted virtual $W$-boson. In particular, the semileptonic processes allow  to probe the kinematic regions of the formfactors that are dominated by the static properties of the baryons.
The recent progress in the lattice quantum chromodynamics gives a hope to determine the properties of the formfactors from the first principles  with the accuracy sufficient for a comparison with precise experimental data~\cite{Detmold:2015aaa}.  Once the hadronic effects are well understood, the SL decays will provide a complementary method to determine Cabbibo--Kobayashi--Maskawa matrix elements~\cite{Cabibbo:2003cu} and to search for beyond the Standard Model effects such as violation of lepton flavour and charge-conjugation--parity  symmetries~\cite{Goudzovski:2022vbt}. In this article, we provide a modular description of the semileptonic decays that can be used to extract properties of the formfactors  in the experiments using spin entangled baryon--antibaryon pairs.

The helicity amplitude method~\cite{Korner:1989qb,Korner:1991ph,Kadeer:2005aq} that is commonly used in the analyses of semileptonic decays allows to express the angular distributions in an efficient and compact way. The complete process is described as a sequence of two-body decays, where each of them is analysed in the rest frame of the subsequent decaying particle. For a semileptonic decay $B_1\to B_2+\ell^-\bar{\nu}_\ell$, the first decay step $B_1\to B_2W^-_{\text{off-shell}}$ is analysed in the $B_1$ rest frame, whereas $W^-_{\text{off-shell}}\to \ell^-\bar{\nu}_\ell$ is analysed in the $W^-_{\text{off-shell}}$ rest frame. The resulting expressions for the differential distributions are compact and can be written in a quasi-factorized form. The formalism also describes joint angular distributions in the semileptonic decays of a spin polarized baryons. 

A novel approach to study strange baryon decays is to use  hyperon--antihyperon pairs from $J/\psi$ resonances produced in electron--positron annihilations~\cite{Faldt:2017kgy}. The complete angular distribution in such processes can be conveniently represented using a product of real-valued matrices  that describe the initial spin-entangled baryon--antibaryon state and chains of two-body weak decays. These matrices can be rearranged to describe many decay scenarios in  the $e^+e^-\to\lam\bar\lam$, $e^+e^-\to \Xi\bar{\Xi}$ and similar processes~\cite{Faldt:2017kgy,Perotti:2018wxm,Adlarson:2019jtw,Salone:2022lpt}. Several high-profile analyses using multidimensional  maximum likelihood fits to angular distributions were performed by the electron--positron collider experiment BESIII~\cite{BESIII:2018cnd,BESIII:2021ypr} using this modular formalism. These multidimensional analyses have demonstrated increased precision of the decay parameters measurements and enabled to observe effects that were averaged out in previous studies, such as a polarization of the hyperon--antihyperon pair from charmonia decays. 
 
The same spin-entangled hyperon--antihyperon system can be used to study semileptonic decays such as $\Lambda\to p e^-\bar\nu_e$ or $\Xi^-\to\Lambda e^-\bar\nu_e$. The processes are relatively rare with the branching fractions (BFs) $8.32(14)\times10^{-4}$ and $5.63(31)\times10^{-4}$, respectively~\cite{Workman:2022ynf}. In the reactions $e^+e^-\to J/\psi\to \Lambda\bar\Lambda$ and $e^+e^-\to J/\psi\to \Xi^-\bar\Xi^+$ the hyperon semileptonic decay is tagged via a common decay of the antihyperon: $\bar\Lambda\to \bar p\pi^+$ and  $\bar\Xi^+\to \bar\Lambda\pi^+$, respectively. The tagging processes involve only charged particles in the final state, therefore their momenta can be precisely determined. This allows one to reconstruct the momentum of the antineutrino in the semileptonic process and to determine the fourmomentum squared of the lepton pair that is needed to study the dynamics of the process. The polarization of the hyperons is given by the angular distributions in their decays, but usually the  polarization of the leptons is not measured. Such double-tag (DT) technique is often used to determine absolute branching fractions in electron--positron collider experiments~\cite{MARK-III:1989dea}. With large  number of collected events in experiments such as BESIII~\cite{BESIII:2021cxx} studies of decay distributions in the semileptonic hyperon decays are possible. A formalism that uses spin correlations and polarization of the produced baryon--antibaryon system is needed to determine  the decay parameters with the best precision.
The purpose of this report is to extend the approach from Refs.~\cite{Perotti:2018wxm,Salone:2022lpt} to include decay matrices representing the three-body semileptonic processes. Our starting point is the helicity formalism for semileptonic decays from Ref.~\cite{Kadeer:2005aq}. We construct a real-valued decay matrix relating the initial and final baryons' spin states, represented by the Pauli matrices. The obtained decay matrix is used to construct the full joint decay distributions of the spin-entangled baryon--antibaryon pair in a modular way.

The paper is organized as follows: in Sec.~\ref{sec:prod} and Sec.~\ref{sec:helamp} we review the formalism to describe baryon--antibaryon production process and semileptonic decays, respectively. In Sec.~\ref{sec:decayM} the main result is derived --- the spin-density matrix of the daughter baryon in the semileptonic decay. Sec.~\ref{sec:jad} presents modular formulas to describe the angular distributions of the semileptonic hyperon decays. Finally in Sec.~\ref{sec:Sens} we collect some numerical results.

\section{Production process}\label{sec:prod}

\begin{figure}
\centering
\includegraphics[width=0.65\columnwidth]{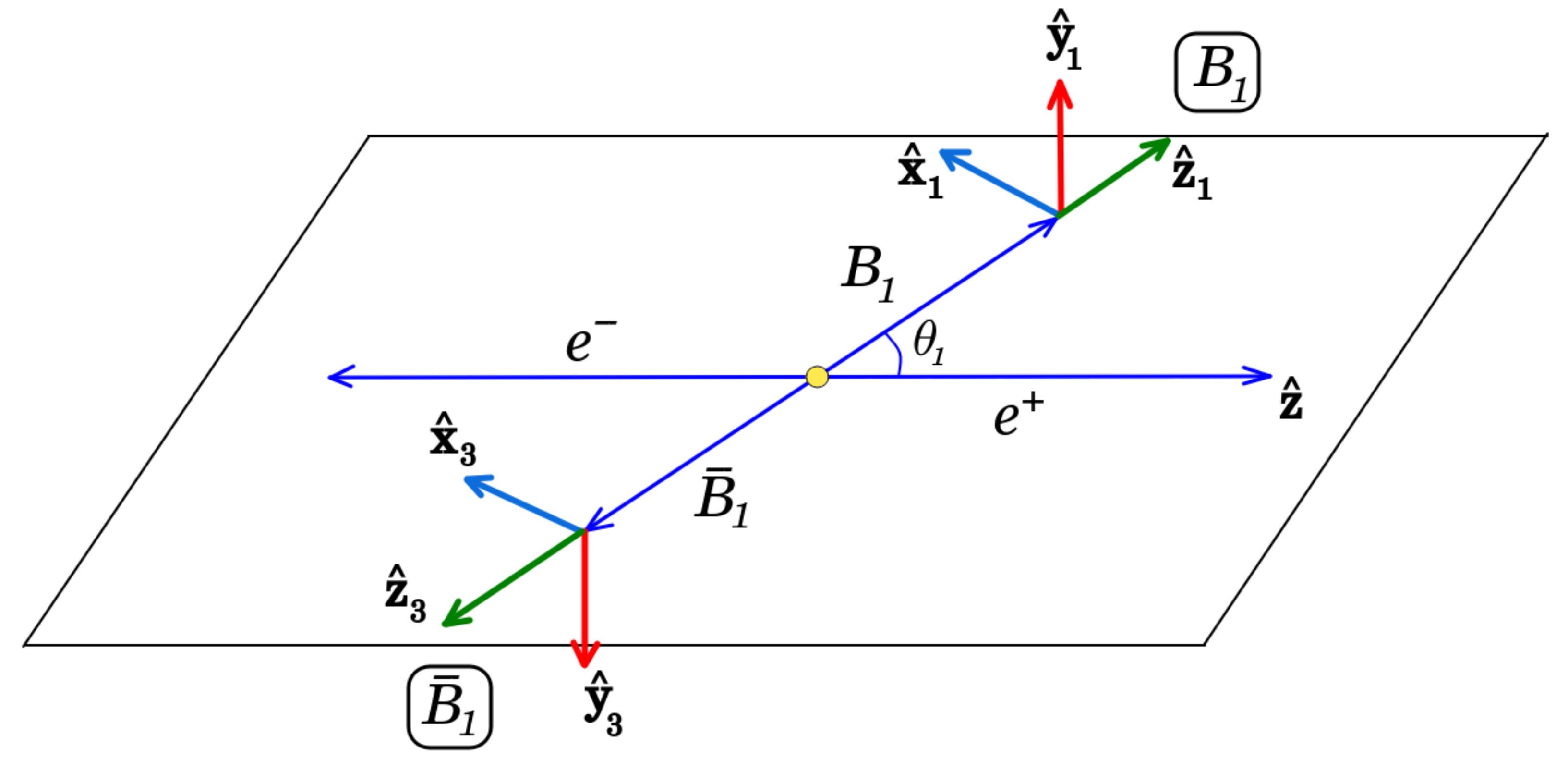}
\caption{Definition of the three coordinate systems used to describe the spin-entangled $B_1\bar B_1$ state. The overall c.m. frame with ${\bf\hat z}$ axis (e.g. for $e^+e^-\to B_1\bar B_1$ it is defined along the positron momentum). The axes in baryon $B_1$ and antibaryon $\bar{B_1}$ rest frames (helicity frames) are denoted $({\bf\hat x}_1,{\bf\hat y}_1,{\bf\hat z}_1)$ and $({\bf\hat x}_3,{\bf\hat y}_3,{\bf\hat z}_3)$, respectively.}
 \label{fig:axes1}
\end{figure}

In general a state of two spin-1/2 particles e.g. a baryon--antibaryon pair $B_1\bar{B}_1$ can be written as~\cite{Perotti:2018wxm}
\begin{equation}
    \rho_{B_1\bar B_1}=\sum_{\mu,\bar{\nu}=0}^3C_{\mu\bar{\nu}}\sigma_{\mu}^{B_1}\otimes\sigma_{\bar{\nu}}^{\bar{B}_1},\label{eq:spincor}
\end{equation}
where a set of four Pauli matrices $\sigma_{\mu}^{B_1} (\sigma_{\bar{\nu}}^{\bar{B}_1})$ acting in the rest frame of a baryon $B_1(\bar{B}_1)$ is used and $C_{\mu\bar{\nu}}$ is a 4$\times$4 real matrix representing polarizations and spin correlations of the baryons. Here we consider mainly baryon--antibaryon systems created in the $e^+e^-\to B_1\bar{B_1}$ process. However, the formalism can be applied for the pairs from decays of (pseudo)scalar or tensor particles such as $\psi(2S),\eta_c,\chi_{c0},\chi_{c2}\to  B_1\bar{B_1}$ or in a fact to any pair of spin-$1/2$ particles (for example baryon--baryon, muon--antimuon and others).
The spin matrices $\sigma_{\mu}^{B_1}$ and $\sigma_{\bar{\nu}}^{\bar{B}_1}$ are given in the coordinate systems with the axes denoted ${\bf\hat x}_1,{\bf\hat y}_1,{\bf\hat z}_1$ and ${\bf\hat x}_3,{\bf\hat y}_3,{\bf\hat z}_3$ as shown in Fig.~\ref{fig:axes1}. The directions of the two right-handed coordinate systems are related as $({{\bf\hat x}_3,{\bf\hat y}_3},{\bf\hat z}_3)=({\bf\hat x}_1,-{\bf\hat y}_1,-{\bf\hat z}_1)$.
 The spin correlation matrix $C_{\mu\bar{\nu}}$ for the reaction $e^+e^-\to B_1\bar{B}_1$ depends in the lowest order on two parameters, $\alpha_{\psi}\in[-1,1]$ and $\Delta\Phi\in[-\pi,\pi)$. The elements of the $C_{\mu\bar{\nu}}$ matrix are functions of the baryon $B_1$  production angle $\theta_1$ in the electron--positron center-of-momentum (c.m.) system. The matrix for the single photon annihilation  of unpolarized electrons and positrons is~\cite{Perotti:2018wxm}:
\begin{equation}
{ C}_{\mu\bar\nu}\propto\begin{pmatrix}
1+\alpha_{\psi}\cos^2\theta_1 & 0 & \beta_{\psi}\sin\theta_1\cos\theta_1 & 0\\
0 & \sin^2\theta_1 & 0 & \gamma_{\psi}\sin\theta_1\cos\theta_1 \\
-\beta_{\psi}\sin\theta_1\cos\theta_1 & 0 & \alpha_{\psi}\sin^2\theta_1 & 0 \\
0 & -\gamma_{\psi}\sin\theta_1\cos\theta_1 & 0 & -\alpha_{\psi}-\cos^2\theta_1
\end{pmatrix},  \label{eq:prod}  
\end{equation}
where the parameters $\beta_{\psi}$ and $\gamma_{\psi}$ are expressed via $\alpha_{\psi}$ and $\Delta\Phi$ as $\beta_{\psi}=\sqrt{1-\alpha^2_{\psi}}\sin(\Delta\Phi)$ and $\gamma_{\psi}=\sqrt{1-\alpha^2_{\psi}}\cos(\Delta\Phi)$. We will also use a more general formula from Ref.~\cite{Salone:2022lpt} that describes the annihilation processes with polarized electron beams.

\section{Invariant formfactors}\label{sec:sl-ff}

Let us consider a semileptonic decay of a 1/2$^+$ hyperon $B_1$ into a 1/2$^+$ baryon $B_2$ and an off-shell $W^-$-boson decaying to the lepton pair $l^-\bar\nu_l$ with the momenta and masses denoted as $B_1(p_1,M_1)\to B_2(p_2,M_2)+l^-(p_l,m_l)+\bar\nu_l(p_{\nu},0)$. The matrix elements due to the vector $J^V_{\mu}$ and axial-vector $J^A_{\mu}$ currents in notation from Ref.~\cite{Kadeer:2005aq} are:
\begin{equation}\label{eq:matrixelem_semil}
 \begin{aligned}
 \langle B_2|J^{V}_{\mu}+J^{A}_{\mu}|B_1\rangle 
& = \bar{u}(p_2)\left[\gamma_{\mu}\left(F^V_1(q^2)+F^A_1(q^2)\gamma_5\right)+\frac{i\sigma_{\mu\nu}q^{\nu}}{M_1}\left(F^V_2(q^2)+F^A_2(q^2)\gamma_5\right)\right.\\
&+\left.\frac{q_{\mu}}{M_1}\left(F^V_3(q^2)+F^A_3(q^2)\gamma_5\right)\right]u(p_1) \ ,
 \end{aligned}
\end{equation}
where $q_{\mu}:=(p_1-p_2)_{\mu}=(p_l+p_\nu)_\mu$ is the fourmomentum transfer. The fourmomentum squared $q^2$ ranges from $m_l^2$ to $(M_1-M_2)^2$. The formfactors $F_{1,2,3}^{V,A}(q^2)$ are complex functions of $q^2$ that describe hadronic effects in the transition.  Neglecting  possible CP-odd weak phases, the  corresponding formfactors are the same for the $(l^-,\bar{\nu}_l)$ and $(l^+,\nu_l)$ transitions. 
To fully determine the hadronic part of a semileptonic decay, the six involved formfactors should be extracted as a function of $q^2$. The formfactors  are usually parameterized by the axial-vector to vector $g_{av}$ coupling, the weak-magnetism  $g_w$ coupling and the pseudoscalar $g_{av3}$ coupling. They are obtained by normalizing to $F_1^V(0)$:
\begin{equation}\label{eq:g_q2}
 g_{av}=\frac{F_1^A(0)}{F_1^V(0)}\ , \qquad g_w=\frac{F_2^V(0)}{F_1^V(0)}\ , \qquad
 g_{av3}=\frac{F_3^A(0)}{F_1^V(0)}\ .
\end{equation}
For experiments with a limited number of events, the $q^2$-dependence of the formfactors is assumed using a model. The standard approach is to include one or more poles of the mesons that have the correct quantum numbers to mix with the $W$ boson and have the masses close to the $q^2$ range in the decay. Traditionally one pole is explicitly included together with  an effective contribution from other poles~\cite{HFLAV:2019otj}
such as in the  Becirevic--Kaidalov (BK)~\cite{Becirevic:1999kt} parameterization:
\begin{equation}\label{eq:beckai}
    F_i(q^2)=\frac{F_i(0)}{1-\frac{q^2}{M^2}}\frac{1}{1-\alpha_{\text{BK}}\frac{q^2}{M^2}}\ ,
\end{equation}
where the dominant pole mass $M$ is outside the kinematic region and the parameter $\alpha_{\text{BK}}$ represent an effective contribution from the meson poles with higher messes. Here the case $\alpha_{\text{BK}}=0$ represents the dominant pole contribution. This parameterization gives real-valued formfactors. If more data is available, one or more extra parameters can be added to describe the $q^2$ distribution. In the hyperon decays the range of $q^2\le(M_1-M_2)^2$ is limited and in the first order can completely neglect the $q^2$ dependence using the values of the couplings at the $q^2=0$ point. A better approximation is to include an effective-range parameter $r_i$ that represent linear dependence on $q^2$:
\begin{equation}
    F_i(q^2)=F_i(0)\left[1+r_iq^2+...\right]\label{eq:g_q2r}\ .
\end{equation}
For example, using~\eqref{eq:beckai} the effective-range parameter is $r_i=(1+\alpha_\text{BK})/M^2$.
The main take-away message from the above discussion is that, for practical purposes, the $q^2$ dependence of an SL formfactor can be represented by one or two parameters. In experiments, these parameters can be determined from the observed distributions. The optimal method for such parametric estimation is the maximum likelihood method using multidimensional unbinned data. We will first construct modular formulas for the angular distributions and then in Sec.~\ref{sec:Sens} discuss the attainable statistical uncertainties for the SL formfactors parameters as the function of the number of observed events.
\section{Helicity amplitudes}\label{sec:helamp}

\begin{figure}
    \centering
    \begin{tikzpicture}
        \draw[<-, ultra thick] (1,0.5) node[anchor=east] {$\hat{x}_1$} -- (2,0);
        \draw[->, ultra thick] (2,0) -- (1.5,1.5) node[anchor=north east] {$\hat{y}_1$};
        \draw[->, ultra thick] (2,0) -- (3.3,0.7) node[anchor=west] {$\hat{z}_1$};
        \draw[->, ultra thick, Orange] (4,2) -- (3,2) node[anchor=east] {$\hat{x}_2$};
        \draw[->, ultra thick, Orange] (4,2) -- (4,3) node[anchor=north east] {$\hat{y}_2$};
        \draw[->, ultra thick, Orange] (4,2) -- (4.5,2.5) node[anchor=north west] {$\hat{z}_2$};
        \draw[->, ultra thick, Cyan] (1,-1) -- (0,-2) node[anchor=north] {$\hat{z}_W$};
        \draw[-,  ] (1,-1) -- (4,2);
        \draw[->, ultra thick, Cyan] (1,-1) -- (-1,-1) node[anchor=north east] {$\hat{x}_W$};
        \draw[->, ultra thick, Cyan] (1,-1) -- (1,-3) node[anchor=north west] {$\hat{y}_W$};
        \draw[<->, ultra thick, Green] (-1,-2) -- (3,0);
        \draw[Green] (0,-1.5) arc (190:211:1);
        \node[Green] at (-0.5,-2) {$\theta_l$};
    \end{tikzpicture}
    \caption{Definition of the three coordinate systems used to describe the semileptonic decay $B_1\to B_2 + W^-_{\text{off-shell}}$. The axes in the $B_1$, $B_2$ and $W^-_{\text{off-shell}}$ rest frames (helicity frames: $\mathbb{R}_1$, $\mathbb{R}_2$ and $\mathbb{R}_W$) are denoted $({\bf\hat x}_1,{\bf\hat y}_1,{\bf\hat z}_1)$, $({\bf\hat x}_2,{\bf\hat y}_2,{\bf\hat z}_2)$ and $({\bf\hat x}_W,{\bf\hat y}_W,{\bf\hat z}_W)$, respectively. \label{fig:hel2}}
\end{figure}
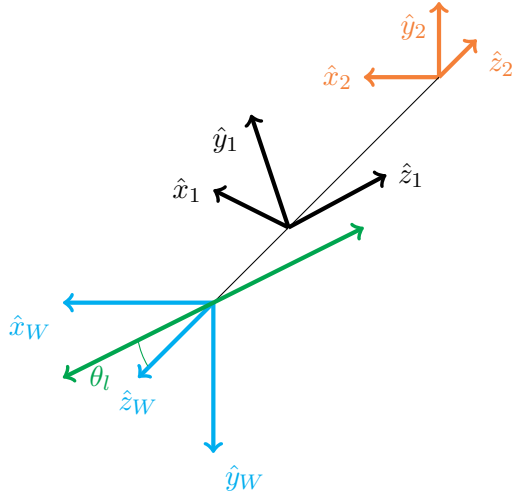

We will describe the $B_1\to B_2+W^-_{\text{off-shell}}$ process using three coordinate systems attached to the three involved particles.  In the baryon $B_1$ rest frame 
$\mathbb{R}_1$, with the $({\bf\hat x}_1,{\bf\hat y}_1,{\bf\hat z}_1)$ Cartesian coordinate system shown in Fig.~\ref{fig:axes1}, the  $B_1$-spin projection on the quantisation axis $\hat{\bf z}_1$ is $\kappa=\pm1/2$. The daughter baryon $B_2$ is emitted in the direction given by the spherical coordinates $\theta_2,\phi_2$ in $\mathbb{R}_1$ and the $B_2$-helicity is $\lambda_2=\pm1/2$. The off-shell $W^-$ boson is emitted in the direction $\theta_W=\pi-\theta_2$, $\phi_W=\pi+\phi_2$ in the  $\mathbb{R}_1$ frame. It has helicity $\ulambda_W=\{t,-1,0,+1\}$ where the time component, $\ulambda_W=t$, corresponds to $J_W=0$ and the remaining three components to $J_W=1$. Therefore, $\ulambda_W$ uniquely defines both spin $J_W$ and helicity  $\lambda_W$ as $J_W(\ulambda_W)=\{0,1,1,1\}$ and $\lambda_W(\ulambda_W)=\{0,-1,0,1\}$, respectively.
The fourmomentum vector of the off-shell $W^-$ is 
$q_{\mu}=\left(q_0,p\sin\theta_W\cos\phi_W ,p\sin\theta_W \sin\phi_W,p\cos\theta_W\right)$ in the $\mathbb{R}_1$ system. The energy $q_0$ of the off-shell $W^-$ boson and the magnitude of the three-momentum $p$ are the following functions of the $q^2$ invariant
\begin{equation}\label{eq:q0}
 q_0(q^2) = \frac{1}{2M_1}(M^2_1-M^2_2+q^2)
\end{equation}
and
\begin{equation}\label{eq:momentum_b2}
 p(q^2)=|{\bf p}_2|=\frac{1}{2M_1}\sqrt{Q_+Q_-},
\end{equation}
where
\begin{equation}\label{eq:Qpm}
 Q_{\pm}=(M_1\pm M_2)^2-q^2.
\end{equation}
The spin direction and subsequent decays of the baryon $B_2$ and boson $W^-_{\text{off-shell}}$ are described in two helicity systems denoted $\mathbb{R}_2$ and $\mathbb{R}_W$, respectively. 
The helicity frame $\mathbb{R}_2$ is obtained by performing three active rotations:  (a) around the $\hat{\bf z}_1$--axis by $-\phi_2$; (b) a rotation around the new $\hat{\bf y}$--axis by $-\theta_2$; (c) a rotation around the $\hat{\bf z}_2$-axis by $+\chi_2$, see Fig.~\ref{fig:hel2} \cite{Jacob:1959at}. The first two rotations are sufficient to align
${\bf p}_2$ with the $z$-axis and such two-rotations prescription is used e.g. in Ref.~\cite{Perotti:2018wxm}. Here we allow for an additional rotation that can be e.g. used to bring the momenta ${\bf p}_2$, ${\bf p}_l$ and ${\bf p}_\nu$ to one plane. Initially, we consider the angle $\chi_2$  of this rotation as an arbitrary parameter. The combined (a)--(c) three-dimensional rotation is given by  the product of three axial rotations ${\cal R}(\chi_2,-\theta_2,-\phi_2)=R_z(\chi_2)R_y(-\theta_2)R_z(-\phi_2)$.
Subsequently, one then boosts to the $B_2$ rest frame. The $\mathbb{R}_W$ frame is defined using the same procedure with the rotation matrix ${\cal R}(\chi_W,-\theta_W,-\phi_W)$ and the subsequent boost to the $W^-_{\text{off-shell}}$ rest frame. Since the $W^-_{\text{off-shell}}$ direction is opposite to $B_2$ in $\mathbb{R}_1$, one has $\phi_W=\pi+\phi_2$ and $\theta_W=\pi-\theta_2$. In order to assure that the coordinate systems in $\mathbb{R}_2$ and $\mathbb{R}_W$ are related as $({\bf\hat x}_2,{\bf\hat y}_2,{\bf\hat z}_2)=({\bf\hat x}_W,-{\bf\hat y}_W,-{\bf\hat z}_W)$ we set $\chi_W=-\chi_2$.

The matching transition amplitude between $B_1$ and the two daughter particles expressed using the defined above helicity frames is~\cite{Jacob:1959at,Perotti:2018wxm}:
\begin{align}
\langle\Omega_2,\lambda_2,\ulambda_W|S|J\!=\!1/2,\kappa\rangle &=\sqrt{\frac{2J+1}{8\pi^2}}\braket{ \lambda_2,\ulambda_W|S|J\!=\!1/2,\kappa}\mathcal{D}^{1/2\ast}_{\kappa,\lambda_2-\lambda_W}(\Omega_2)\nonumber\\
&=\frac{1}{2\pi}H_{\lambda_2,\ulambda_W}(q^2)\mathcal{D}^{1/2\ast}_{\kappa,\lambda_2-\lambda_W}(\Omega_2)\label{eq:helHad}\ , 
\end{align}
where $\mathcal{D}^{J}_{m_1,m_2}(\Omega_2):=\mathcal{D}^{J}_{m_1,m_2}(\phi_2,\theta_2,-\chi_2)$ is the Wigner rotation matrix, where the convention  $\mathcal{D}^{J}_{m_1,m_2}(\phi,\theta,\chi)=e^{-im_1\phi-im_2\chi}\mathcal{D}^{J}_{m_1,m_2}(0,\theta,0)=e^{-im_1\phi-im_2\chi}d^{J}_{m_1,m_2}(\theta)$ is used (see Appendix~\ref{app:pauli}). The order and the signs of the angles $\Omega_2=\{\phi_2,\theta_2,-\chi_2\}$  in the Wigner functions are opposite to the used in the rotations to define the helicity reference frames. In addition, the normalization factor is different since we allow for three independent rotation angles. 
 The helicity  amplitudes $H_{\lambda_2,\ulambda_W}(q^2)$ are functions of $q^2$ and depend on the helicities of the daughter particles.  The vector and axial-vector helicity amplitudes $H_{\lambda_2,\ulambda_W}=H^V_{\lambda_2,\ulambda_W}+H^A_{\lambda_2,\ulambda_W}$ are related to the invariant formfactors in the following way:
\begin{equation}\label{eq:helamp_semil}
\begin{aligned}
 H^V_{\frac{1}{2}t} & = \frac{\sqrt{Q_+}}{\sqrt{q^2}}\left[(M_1-M_2)F^V_1+\frac{q^2}{M_1}F^V_3\right],\\
 H^V_{\frac{1}{2}1} & = \sqrt{2Q_-}\left[-F^V_1-\frac{M_1+M_2}{M_1}F^V_2\right],\\
 H^V_{\frac{1}{2}0} & = \frac{\sqrt{Q_-}}{\sqrt{q^2}}\left[(M_1+M_2)F^V_1+\frac{q^2}{M_1}F^V_2\right],\\
 H^A_{\frac{1}{2}t} & = \frac{\sqrt{Q_-}}{\sqrt{q^2}}\left[-(M_1+M_2)F^A_1+\frac{q^2}{M_1}F^A_3\right],\\
 H^A_{\frac{1}{2}1} & = \sqrt{2Q_+}\left[F^A_1-\frac{M_1-M_2}{M_1}F^A_2\right],\\
 H^A_{\frac{1}{2}0} & = \frac{\sqrt{Q_+}}{\sqrt{q^2}}\left[-(M_1-M_2)F^A_1+\frac{q^2}{M_1}F^A_2\right]\ ,
\end{aligned} 
\end{equation}
where the remaining helicity amplitudes are obtained  by applying the parity operator:
\begin{equation}\label{eq:rel_helampl}
 H^V_{-\lambda_2,-\ulambda_W}=H^V_{\lambda_2,\ulambda_W},\qquad H^A_{-\lambda_2,-\ulambda_W}=-H^A_{\lambda_2,\ulambda_W}\ .
\end{equation}

The decay $W^-\to l^-\bar\nu_l$ is described in $\mathbb{R}_W$ where the emission angles of the $l^-$ lepton are $\theta_l$ and $\phi_l$. The value of the lepton momentum in this frame is 
\begin{equation}\label{eq:lepmom}
    |{\bf p}_l|=\frac{q^2-m_l^2}{2q}\ .
\end{equation}
The decay amplitude reads 
\begin{equation}\label{eq:AmpW}
 \braket{\Omega_l,\lambda_l,\lambda_{\nu}|S_l|J_W,q^2,\lambda_W}=\sqrt{\frac{2J_W+1}{4\pi}}(-1)^{J_W}h^l_{\lambda_l\lambda_{\nu}}(q^2)\mathcal{D}^{J_W\ast}_{\lambda_W,\lambda_l-\lambda_{\nu}}(\Omega_l) \ ,  
\end{equation}
where $\Omega_l=\{\phi_l,\theta_l,0\}$.
The helicity amplitudes $h^l_{\lambda_l\lambda_{\nu}}$  for the elementary transition to the final lepton pair can be calculated directly by evaluating the Feynman diagrams. The neutrino helicities are $\lambda_{\nu}=1/2$  and $\lambda_{\nu}=-1/2$ for ($l^-,\bar{\nu}_l$) and ($l^+,\nu_l$), respectively. The moduli squared of $h^l_{\lambda_l\lambda_{\nu}}$ are \cite{Kadeer:2005aq}:
\begin{align}
  \label{eq:nonflip_lep}
    \mathrm{nonflip} (\ulambda_W=\mp1)& :\  |h^l_{\lambda_l=\mp\frac{1}{2},\lambda_{\nu}=\pm\frac{1}{2}}|^2=8\delta(\lambda_l+\lambda_\nu)(q^2-m_l^2),\\
\label{eq:flip_lep}
    \text{flip} (\ulambda_W=0,t)& :\  |h^l_{\lambda_l=\pm\frac{1}{2},\lambda_{\nu}=\pm\frac{1}{2}}|^2=8\delta(\lambda_l-\lambda_\nu)\frac{m_l^2}{2q^2}(q^2-m_l^2),
\end{align}
where  here and in the following the upper and lower signs refer to the configurations $(l^-,\bar{\nu}_l)$ and $(l^+,\nu_l)$, respectively. 

The representations in Eqs.~\eqref{eq:helHad} and~\eqref{eq:AmpW} imply  that the complete amplitude for the  $B_1(\kappa)\to B_2(\lambda_2)$  transition 
reads:
\begin{equation}
\sum_{\ulambda_W}\braket{\Omega_l,\lambda_l,\lambda_{\nu}|S_l|q^2,\ulambda_W}\langle\Omega_2,\lambda_2,\ulambda_W|S|\!1/2,\kappa\rangle \ ,
\end{equation}
where the  ${\ulambda}_W$ sum runs over the four  $W$-boson helicity components $\{t,-1,0,+1\}$. 
An explicit representation of the amplitude with the angular part separated is 
\begin{equation}
\begin{split}
&\sum_{\ulambda_W}(-1)^{J_W}h^l_{\lambda_l\lambda_{\nu}}\mathcal{D}^{J_{W}\ast}_{\lambda_W,\lambda_l-\lambda_{\nu}}(\Omega_l)   H_{\lambda_2,\ulambda_W}\mathcal{D}^{1/2\ast}_{\kappa,\lambda_2-\lambda_W}(\Omega_2)=\\
&\sum_{\ulambda_W}(-1)^{J_W}h^l_{\lambda_l\lambda_{\nu}}{d}^{J_{W}}_{\lambda_W,\lambda_l-\lambda_{\nu}}(\theta_l)   H_{\lambda_2,\ulambda_W}{d}^{1/2}_{\kappa,\lambda_2-\lambda_W}(\theta_2)\exp\left[{i\kappa\phi_2+i\lambda_2\chi_2-i\lambda_W(\chi_2-\phi_l)}\right],
\end{split}
\end{equation}
where the final expression combines all azimuthal-angle rotations in one term. One can consider two options for selecting $\chi_2$  to define the transversal orientation of the $\mathbb{R}_2$ and $\mathbb{R}_W$ helicity frames. The first option is to set $\chi_2=0$ as in Ref.~\cite{Perotti:2018wxm} where the corresponding azimuthal angle of the charged lepton in the  $\mathbb{R}_W$ system is 
$\phi_l^0$. An alternative is to select $\chi_2^{\text{3-b}}$ so that ${\bf\hat x}_2$ is in the decay plane of the semileptonic decay. In this case the momenta of the leptons are in this plane which corresponds to $\phi_l^{\text{3-b}}=0$  and the $\chi_2^{\text{3-b}}=\phi_l^0$ relation holds. 

The amplitude can be rearranged by inserting a complete spin basis for the baryon $B_2$  to represent transition between $B_1(\kappa)$ and $B_2(\lambda_2)$: 
\begin{align}
&\sum_{\lambda'=-1/2}^{1/2}\sum_{\ulambda_W}\braket{\Omega_l,\lambda_l,\lambda_{\nu}|S_l|q^2,\ulambda_W}\braket{\lambda_2,\ulambda_W|\lambda'}\langle\Omega_2,\lambda'|S|\!1/2,\kappa\rangle  \\
&=\frac{1}{2\pi}\sum_{\lambda'=-1/2}^{1/2}\mathcal{D}^{1/2\ast}_{\kappa,\lambda'}(\Omega_2)\left\{\sum_{\ulambda_W}\braket{\Omega_l,\lambda_l,\lambda_{\nu}|S_l|q^2,\ulambda_W}\braket{\lambda_2,\ulambda_W|\lambda'} H_{\lambda_2,\ulambda_W}(q^2) \right\} \\
    &=\frac{1}{2\pi}\sum_{\lambda'=-1/2}^{1/2}\mathcal{D}^{1/2\ast}_{\kappa,\lambda'}(\Omega_2){\cal H}_{\lambda',\lambda_2}(\Omega_l,q^2,\lambda_l,\lambda_{\nu})\label{eq:DHseparate}\ .
\end{align}
Therefore the angular dependence on $\Omega_2$ can be separated in the amplitude of the complete process. 
Since usually experiments do not measure polarization of the leptons, it is useful to consider a tensor that describes the $W^\pm$-boson decay with the lepton helicities summed over: 
\begin{align}
 L_{\ulambda_W,\ulambda'_W} (q^2,\Omega_l)&:=\sum_{\lambda_l=-1/2}^{1/2} \braket{\Omega_l,\lambda_l,\lambda_{\nu}|S_l|,q^2,\ulambda'_W}^*\braket{\Omega_l,\lambda_l,\lambda_{\nu}|S_l|q^2,\ulambda_W}\\
 &=\frac{3}{4\pi}\sum_{\lambda_l=-1/2}^{1/2}|h^l_{\lambda_l\lambda_{\nu}}(q^2)|^2(-1)^{J_W+J_W'}\mathcal{D}^{J_W\ast}_{\lambda_W,\lambda_l-\lambda_{\nu}}(\Omega_l)\mathcal{D}^{J_W'}_{\lambda_W',\lambda_l-\lambda_{\nu}}(\Omega_l)\label{eq:ltensor}\\
 &=\frac{3}{4\pi}e^{i(\lambda_W-\lambda'_W)\phi_l}\sum_{\lambda_l=-1/2}^{1/2}|h^l_{\lambda_l\lambda_{\nu}}(q^2)|^2(-1)^{J_W+J'_W}d^{J_W}_{\lambda_W,\lambda_l-\lambda_{\nu}}(\theta_l)d^{J_W'}_{\lambda_W',\lambda_l-\lambda_{\nu}}(\theta_l)\label{eq:lTensorA}\ . 
\end{align}
The interference contribution from $\ulambda_W=t$ and $\ulambda_W=0$ gives an extra minus sign. We write the tensor as:
\begin{equation}
        L_{\ulambda_W,\ulambda'_W} (q^2,\Omega_l)=\frac{6}{\pi}(q^2-m_l^2)\left[\ell_{\ulambda_W,\ulambda'_W}^\text{nf} (\Omega_l)+\varepsilon\ell_{\ulambda_W,\ulambda'_W}^\text{f} (\Omega_l)\right]\ ,
\end{equation}
where $\varepsilon={m_l^2}/{(2q^2})$.
The hermitian matrix for the nonflip transition reads
\begin{equation}
\ell_{\ulambda_W,\ulambda'_W}^\text{nf} (\Omega_l)=\left(
\begin{array}{cccc}
 0 & 0 & 0 & 0 \\
 0 &\frac{(1\pm\cos\theta_l)^2 }{4} & \mp\frac{e^{-i \phi_l } \sin \theta_l  (1\pm\cos \theta_l)}{2 \sqrt{2}} & \frac14e^{-2 i \phi_l }\sin ^2\theta_l \\
 0 & \mp\frac{e^{i \phi_l } \sin \theta_l  (1\pm\cos \theta_l)}{2 \sqrt{2}} & \frac12\sin ^2\theta_l & \mp\frac{e^{-i \phi_l } \sin \theta_l  (1\mp\cos \theta_l)}{2 \sqrt{2}} \\
 0 & \frac14e^{2 i \phi_l } \sin ^2\!\theta_l & \mp\frac{e^{i \phi_l } \sin \theta_l  (1\mp\cos \theta_l)}{2 \sqrt{2}} & \frac{(1\mp\cos\theta_l)^2 }{4} \\
\end{array}
\right)\ ,
\end{equation}
while for the flip transition 
\begin{equation}
\ell_{\ulambda_W,\ulambda'_W}^\text{f}  (\Omega_l)=  \left(
\begin{array}{cccc}
 1 & -\frac{e^{i\phi_l} \sin\theta_l}{\sqrt{2}} & -\cos\theta_l & \frac{e^{-i\phi_l} \sin\theta_l}{\sqrt{2}} \\
 -\frac{e^{-i\phi_l} \sin\theta_l}{\sqrt{2}} & \frac{\sin ^2\theta_l}{2} & \frac{e^{-i\phi_l} \sin\theta_l \cos\theta_l}{\sqrt{2}} & -\frac{1}{2} e^{-2 i\phi_l} \sin ^2\!\theta_l \\
 -\cos\theta_l & \frac{e^{i\phi_l} \sin\theta_l \cos\theta_l}{\sqrt{2}} & \cos ^2\theta_l & -\frac{e^{-i\phi_l} \sin\theta_l \cos\theta_l}{\sqrt{2}} \\
 \frac{e^{i\phi_l} \sin\theta_l}{\sqrt{2}} & -\frac{1}{2} e^{2 i\phi_l} \sin ^2\!\theta_l & -\frac{e^{i\phi_l} \sin\theta_l \cos\theta_l}{\sqrt{2}} & \frac12\sin ^2\theta_l\\
\end{array}
\right)\ .
\end{equation}

\section{Decay matrix}\label{sec:decayM}

Here, we derive a matrix that relates the spin of the baryon $B_2$ to the spin of the baryon $B_1$ in $B_1\to B_2 \ell\nu_\ell$ where the state of the lepton pair with the summed spin projections is given by the $L_{\ulambda_W,\ulambda'_W} (q^2,\Omega_l)$ tensor in Eq.~\eqref{eq:ltensor}. The transition can be represented by a tensor $T^{\kappa\kappa',\lambda_2\lambda_2'}$ that describes how the initial spin-density matrix  $\rho^{\kappa\kappa'}_1$ of the baryon $B_1$ transforms to the density matrix $\rho^{\lambda_2\lambda_2'}_2$ of the baryon $B_2$:
\begin{equation}
    \rho^{\lambda_2\lambda_2'}_2=T^{\kappa\kappa',\lambda_2\lambda_2'}\rho^{\kappa\kappa'}_1\ .
\end{equation}
Using Eq.~\eqref{eq:helHad} the transition tensor is given as
\begin{align}
T^{\kappa\kappa',\lambda_2\lambda_2'} &= {\frac{1}{4\pi^2}}\sum_{\ulambda_W,\ulambda_W'}H_{\lambda_2\ulambda_W}H^{\ast}_{\lambda_2'\ulambda_W'} \mathcal{D}^{1/2\ast}_{\kappa,\lambda_2-\lambda_W}(\Omega_2)\mathcal{D}^{1/2}_{\kappa',\lambda_2'-\lambda_W'}(\Omega_2) L_{\lambda_W,\lambda'_W} (q^2,\Omega_l)\\
&\equiv \frac{1}{4\pi^2}\sum_{\ulambda_W,\ulambda_W'}T^{\kappa\kappa',\lambda_2\lambda_2'}_{\ulambda_W,\ulambda_W'}(q^2,\Omega_2)L_{\ulambda_W,\ulambda'_W} (q^2,\Omega_l)
\ .
\end{align}
The explicit expression for the phases of the hadronic tensor due to the azimuthal rotations is
\begin{equation}
\begin{split}
T^{\kappa\kappa',\lambda_2\lambda_2'}_{\ulambda_W,\ulambda_W'}(q^2,\Omega_2) 
&=H_{\lambda_2\ulambda_W}H^{\ast}_{\lambda_2'\ulambda_W'} {d}^{1/2}_{\kappa,\lambda_2-\lambda_W}(\theta_2){d}^{1/2}_{\kappa',\lambda_2'-\lambda_W'}(\theta_2)\\
&\times\exp\left[{i\kappa\phi_2+i\lambda_2\chi_2-i\lambda_W\chi_2}\right]\\ 
&\times\exp\left[{-i\kappa'\phi_2-i\lambda_2'\chi_2+i\lambda_W'\chi_2}\right]\ ,
\end{split}
\end{equation}
where we use the generic case with $\Omega_2=\{\phi_2,\theta_2,\chi_2\}$ and $\Omega_l=\{\phi_l,\theta_l,0\}$. The overall phases of the contraction of the above hadronic tensor and the leptonic tensor in Eq.~\eqref{eq:lTensorA} for the two choices of the orientations of the coordinate systems $\mathbb{R}_2$ and $\mathbb{R}_W$ are:
\begin{align}
    (\chi_2=0)\to& \exp\left[{i(\kappa-\kappa')\phi_2+i(\lambda_W-\lambda_W')\phi_l^0}\right],\\
    (\phi_l^{\text{3-b}}=0)\to& \exp\left[{i(\kappa-\kappa')\phi_2+i(\lambda_2-\lambda_2')\chi_2^{\text{3-b}}}\right]\\
    =&\exp\left[{i(\kappa-\kappa')\phi_2+i(\lambda_2-\lambda_2')\phi_l^0}\right]\ .
\end{align}
The two representations are not equivalent but can be written in terms of the tensors evaluated for $\Omega_2^0:=\{\phi_2,\theta_2,0\}$ and $\Omega_l^0=\{0,\theta_l,0\}$ as 

\begin{align}
 T^{\kappa\kappa',\lambda_2\lambda_2'}(\chi_2=0) &=  \frac{1}{{4\pi^2}}\sum_{\ulambda_W,\ulambda_W'}\exp\left[{i(\lambda_W-\lambda_W')\phi_l^0}\right]T^{\kappa\kappa',\lambda_2\lambda_2'}_{\ulambda_W,\ulambda_W'}(\Omega_2^0)L_{\ulambda_W,\ulambda'_W} (\Omega_l^0),\\
 T^{\kappa\kappa',\lambda_2\lambda_2'}(\phi_l^{\text{3-b}}=0) &=  \frac{1}{{4\pi^2}}\exp\left[{i(\lambda_2-\lambda_2')\phi_l^0}\right]\sum_{\ulambda_W,\ulambda_W'}T^{\kappa\kappa',\lambda_2\lambda_2'}_{\ulambda_W,\ulambda_W'}(\Omega_2^0)L_{\ulambda_W,\ulambda'_W} (\Omega_l^0) 
\ .
\end{align}

Instead of the helicities, the transition can be written as in Ref.~\cite{Perotti:2018wxm} using  spin base vectors $\sigma^{B_1}_{\mu}$ and  $\sigma^{B_2}_{\nu}$ in the mother and daughter reference systems $\mathbb R_1$ and  $\mathbb R_2$, respectively. The $4\times4$ matrix $\BB_{\mu\nu}$ describes how the decay process transforms the base Pauli matrices:
\begin{align}
 \sigma_{\mu}^{B_1}\to&{\frac{3(q^2-m_l^2)}{4\pi^3}}\sum^3_{\nu=0}\BB_{\mu\nu}\sigma^{B_2}_{\nu}\label{eq:sigma_matrix_L_semil}\ .
\end{align}
The real coefficients $\BB_{\mu\nu}$  can be  obtained by  inserting Pauli $\sigma_\mu$ matrices for the mother and the daughter baryons in the expression for the tensor  $T^{\kappa\kappa',\lambda_2\lambda_2'}$:
\begin{equation}
\begin{split}
\BB_{\mu\nu}:= & \frac{2\pi^3}{{3(q^2-m_l^2)}}\sum_{\lambda_2,\lambda_2'=-1/2}^{1/2}\sum_{\kappa,\kappa'=-1/2}^{1/2}T^{\kappa\kappa',\lambda_2\lambda_2'}{\sigma_{\mu}^{\kappa,\kappa'}\sigma_{\nu}^{\lambda_2',\lambda_2} }\ .
\end{split}
\end{equation}
However, as we show in Appendix~\ref{sec:DerivRBb} the coefficients can be represented as
\begin{align}
 \BB_{\mu\nu}&=\sum_{\kappa=0}^3{\cal R}_{\mu\kappa}^{(4)}( \Omega_2) b_{\kappa\nu}(q^2,\Omega_l)\label{eq:BRb} \ ,
\end{align}
where ${\cal R}_{\mu\kappa}^{(4)}( \Omega_2)$ is  the $4\times4$ space-like rotation matrix obtained as the direct sum of identity and 3D rotation $ {\cal R}( \Omega_2)$:
   ${\cal R}^{(4)}( \Omega_2)= {\rm diag}\left(1,  {\cal R}( \Omega_2)\right)$.
The argument $\Omega_2=\{\phi_2,\theta_2,-\chi_2\}$ assures that the rotation is the inverse of the rotation ${\cal R}(\{\chi_2,-\theta_2,-\phi_2\})$  that was used to define the helicity frame $\mathbb{R}_2$.
The coefficients $b_{\mu\nu}$ correspond to the $B_1\to B_2$ transition where the orientations of the axes of the reference systems are aligned $\Omega_2=\{0,0,0\}$. They can be  obtained by  inserting Pauli $\sigma_\mu$ matrices for the mother and the daughter baryons in the expression for the tensor  $T^{\kappa\kappa',\lambda_2\lambda_2'}$ with $\Omega_2$ set to $ \{0,0,0\}$ what implies replacement $\mathcal{D}^{1/2}_{m_1,m_2}(\{0,0,0\})=\delta(m_1-m_2)$:
\begin{equation}
\begin{split}
b_{\mu\nu}:= & \frac{\pi}{{6(q^2-m_l^2)}} \sum_{\ulambda_W,\ulambda_W'}\sum_{\lambda_2,\lambda_2'=-1/2}^{1/2}H_{\lambda_2\ulambda_W}H^{\ast}_{\lambda_2'\ulambda_W'}\underbrace{\sigma_{\mu}^{\lambda_2-\lambda_W,\lambda_2'-\lambda_W'}\sigma_{\nu}^{\lambda_2',\lambda_2} L_{\ulambda_W,\ulambda'_W} (q^2,\Omega_l)}_{{\cal T}^{\ulambda_W,\ulambda_W',\lambda_2,\lambda_2'}_{\mu\nu}}\ ,
\end{split}\label{eq:bmunu_semil}
\end{equation}
\begin{equation}\label{eq:b_L_semil}
\begin{aligned}
b_{\mu\nu}= & \sum_{\ulambda_W,\ulambda_W'}\sum_{\lambda_2,\lambda_2'=-1/2}^{1/2}H_{\lambda_2\ulambda_W}H^{\ast}_{\lambda_2'\ulambda_W'}{\cal T}^{\ulambda_W,\ulambda_W',\lambda_2,\lambda_2'}_{\mu\nu}\\
 =&\sum_{\ulambda_W}\sum_{\lambda_2=-1/2}^{1/2}\left\{\vphantom{\sum_{\lambda_2=-1/2}^{1/2}}
  \left|H_{\lambda_2\ulambda_W}\right|^2 
  {\cal T}^{\ulambda_W,\ulambda_W,\lambda_2,\lambda_2}_{\mu\nu}\right.\\
 +&\left. 2\sum_{
 \ulambda_W'<\ulambda_W}\sum_{\lambda_2'<\lambda_2}\left[\Re(H_{\lambda_2\ulambda_W}H^{\ast}_{\lambda_2'\ulambda_W'}) \Re{\cal T}^{\ulambda_W,\ulambda_W',\lambda_2,\lambda_2'}_{\mu\nu}
 +
 \Im(H_{\lambda_2\ulambda_W}H^{\ast}_{\lambda_2'\ulambda_W'}) \Im{\cal T}^{\ulambda_W,\ulambda_W',\lambda_2,\lambda_2'}_{\mu\nu} \right]
\right\}.
\end{aligned}
\end{equation}
The last form involves only real valued tensors ${\cal T}^{\ulambda_W,\ulambda_W,\lambda_2,\lambda_2}_{\mu\nu}$, $\Re{\cal T}^{\ulambda_W,\ulambda_W',\lambda_2,\lambda_2'}_{\mu\nu}$ and $\Im{\cal T}^{\ulambda_W,\ulambda_W',\lambda_2,\lambda_2'}_{\mu\nu}$. The hadronic part is encoded in the  real-valued functions of $q^2$:
$\left|H_{\lambda_2\ulambda_W}\right|^2$,
$\Re(H_{\lambda_2\ulambda_W}H^{\ast}_{\lambda_2'\ulambda_W'})$ and 
$\Im(H_{\lambda_2\ulambda_W}H^{\ast}_{\lambda_2'\ulambda_W'})$, where $\ulambda_W'<\ulambda_W$ and $\lambda_2'<\lambda_2$. Moreover, the formfactors  $H_{-\frac12 1}=H_{\frac12 -1}=0$ reducing number of the functions.

We will represent the $b_{\mu\nu}$ matrix as the sum of the non-flip and flip contributions
$b_{\mu\nu}=b^\mathrm{nf}_{\mu\nu}+\varepsilon b^\mathrm{f}_{\mu\nu}$. 
The cross-section term is written as $b_{00}=b_{00}^\textrm{nf}+\varepsilon b_{00}^\textrm{f}$ where 
\begin{align}\label{eq:norm_L_semil}
    b_{00}^\textrm{nf}&=\frac{1}{4}(1\mp\cos\theta_l)^2|H_{\frac{1}{2}1}|^2+\frac{1}{4}(1\pm\cos\theta_l)^2|H_{-\frac{1}{2}-1}|^2+\frac{1}{2}\sin^2\theta_l(|H_{-\frac{1}{2}0}|^2+|H_{\frac{1}{2}0}|^2),\\
    b_{00}^\textrm{f}&=|H_{\frac{1}{2}t}|^2+|H_{-\frac{1}{2}t}|^2+\frac12\sin^2\theta_l(|H_{\frac{1}{2}1}|^2+|H_{-\frac{1}{2}-1}|^2)+ \cos^2\theta_l(|H_{\frac{1}{2}0}|^2+|H_{-\frac{1}{2}0}|^2)\\
    &-2\cos\theta_l\Re(H^\ast_{\frac{1}{2}0}H_{\frac{1}{2}t}+H^\ast_{-\frac{1}{2}0}H_{-\frac{1}{2}t}) \ ,\nonumber
\end{align}
define the angular distributions for the decay of unpolarized baryon $B_1$ when the spins of all final particles are summed over. The differential decay rate  is obtained by multiplying by the kinematic and spinor normalization factors that depend on $q^2$
\begin{align}\label{eq:dg}
    {\dd\Gamma}&= \frac{G_F^2}{(2\pi)^5}|V_{us}|^2\frac{|{\bf p}_l| |{\bf p}_2|}{16 M_1^2} (q^2-m_l^2)b_{00}{\dd q \dd\Omega_2 \dd\Omega_l}\\
    &= {G_F^2}|V_{us}|^2{V_{Ph}(q^2)} (q^2-m_l^2)b_{00}{\dd q \dd\Omega_2 \dd\Omega_l}\ ,
\end{align}
where $V_{Ph}(q^2)= {(2\pi)^{-5}}{(4 M_1)^{-2}}{|{\bf p}_l| |{\bf p}_2|}$ is the three-body phase space density factor~\cite{Workman:2022ynf}. The momenta $|{\bf p}_2|$ and $|{\bf p}_l|$ of the baryon $B_2$ and the lepton are given in Eqs.~\eqref{eq:momentum_b2} and~\eqref{eq:lepmom}, respectively.

The first row of the $b_{0i}$ matrix, where $i=1,2,3\ (x,y,z)$, gives the polarization vector ${\bf P}=(P_x,P_y,P_z)$ of the baryon $B_2$ in the reference frame  $\mathbb R_2$ corresponding to the decay of unpolarized baryon $B_1$. These elements are:
\begin{equation}\label{eq:b0i}
\begin{aligned}
  b_{01} & =-\Re(\mathcal{I}_{01})\cos\phi_l+\Im(\mathcal{I}_{01})\sin\phi_l=P_x b_{00}, \\
 b_{02} & =\phantom{-}\Re(\mathcal{I}_{01})\sin\phi_l+\Im(\mathcal{I}_{01})\cos\phi_l=P_y b_{00}, \\
 b_{03} & =\phantom{-}b_{03}^\textrm{nf}+\varepsilon b_{03}^\textrm{f}=P_z b_{00},\\
\end{aligned}
\end{equation}
where $\mathcal{I}_{\mu\nu}$ are complex. We use notation $\mathcal{I}_{\mu\nu}=\mathcal{I}_{\mu\nu}^\textrm{nf}+\varepsilon \mathcal{I}_{\mu\nu}^\textrm{f}$ and 
\begin{equation}\label{eq:decayparam_L_semil}
\begin{aligned}
    \mathcal{I}_{01}^\textrm{nf} & =\pm\frac{1}{\sqrt{2}}\sin\theta_l\left[(1\pm\cos\theta_l)H^\ast_{-\frac{1}{2}-1}H_{\frac{1}{2}0}+(1\mp\cos\theta_l)H^{\ast}_{-\frac{1}{2}0}H_{\frac{1}{2}1}\right],\\
    \mathcal{I}_{01}^\textrm{f} & =\sqrt{2}\sin\theta_l\left[(H^\ast_{-\frac{1}{2}-1}H_{\frac{1}{2}t}-H^\ast_{-\frac{1}{2}t}H_{\frac{1}{2}1})
    +\cos\theta_l (H^\ast_{-\frac{1}{2}0}H_{\frac{1}{2}1}-H^\ast_{-\frac{1}{2}-1}H_{\frac{1}{2}0})\right]\ ,\\
    b_{03}^\textrm{nf} & =\frac{1}{4}(1\mp\cos\theta_l)^2|H_{\frac{1}{2}1}|^2-\frac{1}{4}(1\pm\cos\theta_l)^2|H_{-\frac{1}{2}-1}|^2-\frac{1}{2}\sin^2\theta_l(|H_{-\frac{1}{2}0}|^2-|H_{\frac{1}{2}0}|^2),\\
    b_{03}^\textrm{f} & =|H_{\frac{1}{2}t}|^2-|H_{-\frac{1}{2}t}|^2+\frac{1}{2}\sin^2\theta_l(|H_{\frac{1}{2}1}|^2-|H_{-\frac{1}{2}-1}|^2)-\cos^2\theta_l(|H_{-\frac{1}{2}0}|^2-|H_{\frac{1}{2}0}|^2)\\
    & -2\cos\theta_l\Re(H^{\ast}_{\frac{1}{2}0}H_{\frac{1}{2}t}-H^{\ast}_{-\frac{1}{2}0}H_{-\frac{1}{2}t})\ .\\
\end{aligned}    
\end{equation}
The first column  $b_{i0}$ of the matrix corresponds to the decay of the spin polarized 
baryon $B_1$. The element $b_{30}=b_{30}^\mathrm{nf}+\varepsilon b_{30}^\mathrm{f}$
is:
\begin{equation}\label{eq:decparam_b30}
\begin{aligned}
    b_{30}^\textrm{nf} & =\frac{1}{4}(1\pm\cos\theta_l)^2|H_{-\frac{1}{2}-1}|^2-\frac{1}{4}(1\mp\cos\theta_l)^2|H_{\frac{1}{2}1}|^2-\frac{1}{2}\sin^2\theta_l(|H_{-\frac{1}{2}0}|^2-|H_{\frac{1}{2}0}|^2)\ ,\\
    b_{30}^\textrm{f} & =|H_{\frac{1}{2}t}|^2-|H_{-\frac{1}{2}t}|^2-\frac{1}{2}\sin^2\theta_l(|H_{\frac{1}{2}1}|^2-|H_{-\frac{1}{2}-1}|^2)-\cos^2\theta_l(|H_{-\frac{1}{2}0}|^2-|H_{\frac{1}{2}0}|^2)\\
    & -2\cos\theta_l\Re(H^{\ast}_{\frac{1}{2}0}H_{\frac{1}{2}t}-H^{\ast}_{-\frac{1}{2}0}H_{-\frac{1}{2}t}) \ .\\
\end{aligned}    
\end{equation}
The elements $b_{10}$ and $b_{20}$ are
\begin{equation}\label{eq:bi0}
\begin{aligned}
  b_{10}&=-\cos\phi_l \Re(\mathcal{I}_{10})+\sin\phi_l  \Im(\mathcal{I}_{10})\ ,\\
 b_{20}&=\phantom{-} \sin\phi_l\Re(\mathcal{I}_{10})+\cos\phi_l\Im(\mathcal{I}_{10})\ ,
\end{aligned}
\end{equation}
where
\begin{equation}\label{eq:decparam_bi0}
\begin{aligned}
   \mathcal{I}_{10}^\text{nf}& =\pm\frac{1}{\sqrt{2}}\sin\theta_l \left[(1\pm\cos\theta_l)H^{\ast}_{-\frac{1}{2}-1}H_{-\frac{1}{2}0}+(1\mp\cos\theta_l)H^{\ast}_{\frac{1}{2}0}H_{\frac{1}{2}1}\right],\\
    \mathcal{I}_{10}^\text{f} & ={\sqrt{2}}\sin\theta_l \left[(H^\ast_{-\frac{1}{2}-1}H_{-\frac{1}{2}t}
     -H^\ast_{\frac{1}{2}t}H_{\frac{1}{2}1})
    +\cos\theta_l(H^\ast_{\frac{1}{2}0}H_{\frac{1}{2}1}-H^\ast_{-\frac{1}{2}-1}H_{-\frac{1}{2}0})\right]\ .
\end{aligned}    
\end{equation}
The decay plane representation which requires three rotation angles for baryon  $B_2$ gives simple formulas for the remaining terms of the decay matrix. 
The terms of the non-flip contributions for the aligned (with $\phi_l=0$) decay matrix $b^\textrm{nf}_{\mu\nu}$ are: 
\begin{equation}\label{eq:bnf}
    b_{\mu\nu}^\textrm{nf}=\left(
\begin{array}{cccc}
b_{00}^\textrm{nf} &-\Re(\mathcal{I}_{01}^\textrm{nf}) & \Im(\mathcal{I}_{10}^\textrm{nf}) & b_{03}^\textrm{nf} \\
 -\Re(\mathcal{I}_{10}^\textrm{nf})&\Re(\mathcal{E}_{00}^\textrm{nf}+\mathcal{E}_{11}^\textrm{nf}) & -\Im(\mathcal{E}_{00}^\textrm{nf}+\mathcal{E}_{11}^\textrm{nf})& \phantom{-}\Re(\mathcal{I}_{13}^{\textrm{nf}})\\
 \phantom{-}\Im(\mathcal{I}_{10}^\textrm{nf})& \Im(\mathcal{E}_{00}^\textrm{nf}-\mathcal{E}_{11}^\textrm{nf})&\phantom{-}\Re(\mathcal{E}_{00}^\textrm{nf}-\mathcal{E}_{11}^\textrm{nf}) &-\Im(\mathcal{I}_{13}^{\textrm{nf}}) \\
b_{30}^\textrm{nf} & -\Re(\mathcal{I}_{31}^{\textrm{nf}}) &\phantom{-}\Im(\mathcal{I}_{31}^{\textrm{nf}})& b_{33}^\textrm{nf} \\
\end{array}
\right) \ ,  
\end{equation}
where
\begin{equation}
\begin{aligned}
 b^\textrm{nf}_{33} & =\frac{1}{2}\sin^2\theta_l(|H_{-\frac{1}{2}0}|^2+|H_{\frac{1}{2}0}|^2)-\frac{1}{4}(1\mp\cos\theta_l)^2|H_{\frac{1}{2}1}|^2-\frac{1}{4}(1\pm\cos\theta_l)^2|H_{-\frac{1}{2}-1}|^2 \\
\end{aligned}
\end{equation}
and
\begin{equation}
\begin{aligned}
    \mathcal{I}_{13}^{\textrm{nf}} & =\pm\frac{1}{\sqrt{2}}\sin\theta_l\left\{(1\pm\cos\theta_l)H^{\ast}_{-\frac{1}{2}-1}H_{-\frac{1}{2}0}-(1\mp\cos\theta_l)H^{\ast}_{\frac{1}{2}0}H_{\frac{1}{2}1}\right\},\\
    \mathcal{I}_{31}^{\textrm{nf}} & =\pm\frac{1}{\sqrt{2}}\sin\theta_l\left\{(1\pm\cos\theta_l)H^{\ast}_{-\frac{1}{2}-1}H_{\frac{1}{2}0}-(1\mp\cos\theta_l)H^{\ast}_{-\frac{1}{2}0}H_{\frac{1}{2}1}\right\},\\
    \mathcal{E}_{00}^\textrm{nf} & =\sin^2\theta_l H^{\ast}_{-\frac{1}{2}0}H_{\frac{1}{2}0},\\
    \mathcal{E}_{11}^\textrm{nf} & =\frac{1}{2}\sin^2\theta_l H^{\ast}_{-\frac{1}{2}-1}H_{\frac{1}{2}1}.\\
\end{aligned}    
\end{equation}
The terms of the flip contributions for the aligned decay matrix $b^\textrm{f}_{\mu\nu}$ are: 
\begin{equation}\label{eq:flip}
    b_{\mu\nu}^\textrm{f}=\left(
\begin{array}{cccc}
b_{00}^\textrm{f} &-\Re(\mathcal{I}_{01}^\textrm{f}) & \Im(\mathcal{I}_{10}^\textrm{f}) & b_{03}^\textrm{f} \\
 -\Re(\mathcal{I}_{10}^\textrm{f})&\Re(\mathcal{E}_{00}^\textrm{f}-\mathcal{E}_{11}^\textrm{f}) & -\Im(\mathcal{E}_{00}^\textrm{f}-\mathcal{E}_{11}^\textrm{f})& \phantom{-}\Re(\mathcal{I}_{13}^{\textrm{f}})\\
 \phantom{-}\Im(\mathcal{I}_{10}^\textrm{f})& \Im(\mathcal{E}_{00}^\textrm{f}+\mathcal{E}_{11}^\textrm{f})&\phantom{-}\Re(\mathcal{E}_{00}^\textrm{f}+\mathcal{E}_{11}^\textrm{f}) &-\Im(\mathcal{I}_{13}^{\textrm{f}}) \\
b_{30}^\textrm{f} & -\Re(\mathcal{I}_{31}^{\textrm{f}}) &\phantom{-}\Im(\mathcal{I}_{31}^{\textrm{f}})& b_{33}^\textrm{f} \\
\end{array}
\right) \ ,  
\end{equation}
where
\begin{equation}
\begin{aligned}
 b_{33}^\textrm{f} & =     |H_{-\frac{1}{2}t}|^2+|H_{\frac{1}{2}t}|^2+\cos^2\theta_l(|H_{-\frac{1}{2}0}|^2+|H_{\frac{1}{2}0}|^2)-\frac{1}{2}\sin^2\theta_l(|H_{\frac{1}{2}1}|^2+|H_{-\frac{1}{2}-1}|^2)\\
    &-2\cos\theta_l\Re(H^{\ast}_{\frac{1}{2}0}H_{\frac{1}{2}t}+H^{\ast}_{-\frac{1}{2}0}H_{-\frac{1}{2}t}) \\
\end{aligned}
\end{equation}
and 
\begin{equation}
\begin{aligned}
    \mathcal{I}_{13}^\textrm{f} & =\sqrt{2}\sin\theta_l\left\{H^{\ast}_{-\frac{1}{2}-1}H_{-\frac{1}{2}t}+H^{\ast}_{\frac{1}{2}1}H_{\frac{1}{2}t}-\cos\theta_l (H^{\ast}_{\frac{1}{2}0}H_{\frac{1}{2}1}+H^{\ast}_{-\frac{1}{2}-1}H_{-\frac{1}{2}0})\right\},\\
    \mathcal{I}_{31}^\textrm{f} & =\sqrt{2}\sin\theta_l\left\{H^{\ast}_{-\frac{1}{2}-1}H_{\frac{1}{2}t}+H^{\ast}_{-\frac{1}{2}t}H_{\frac{1}{2}1}-\cos\theta_l (H^{\ast}_{-\frac{1}{2}0}H_{\frac{1}{2}1}+H^{\ast}_{-\frac{1}{2}-1}H_{\frac{1}{2}0})\right\},\\
    \mathcal{E}_{00}^\textrm{f} & =2\left\{H^{\ast}_{-\frac{1}{2}t}H_{\frac{1}{2}t}+\cos^2\theta_l H^{\ast}_{-\frac{1}{2}0}H_{\frac{1}{2}0}-\cos\theta_l( H^{\ast}_{-\frac{1}{2}0}H_{\frac{1}{2}t}+H^{\ast}_{-\frac{1}{2}t}H_{\frac{1}{2}0})\right\},\\
    \mathcal{E}_{11}^\textrm{f} & =\sin^2\theta_l H^{\ast}_{-\frac{1}{2}-1}H_{\frac{1}{2}1}\ .\\
\end{aligned}    
\end{equation}
If the formfactors have no complex phases, meaning the $\mathcal{I}_{\mu\nu}$ terms are real functions, the decay matrix reads as
\begin{equation}
    b_{\mu\nu}=\left(
\begin{array}{cccc}
b_{00} &-\mathcal{I}_{01} & 0 & \mathcal{I}_{03} \\
 -\mathcal{I}_{10}&b_{11} & 0& \mathcal{I}_{13}\\
 0& 0&b_{22} & 0 \\
- \mathcal{I}_{30} & \mathcal{I}_{31} & 0& b_{33} \\
\end{array}
\right) \ .   
\end{equation}
The terms of the $b_{\mu\nu}$ matrix in general form for an arbitrary $\phi_l$ value are given in Appendix~\ref{app:nf}. They should be used if two rotation angle representation as in Ref.~\cite{Perotti:2018wxm} was applied.

\section{Joint angular distributions}\label{sec:jad}

Here we provide examples how to construct modular expressions for the angular distributions of semileptonic decays of baryons. First, using our formalism, we rewrite the results from Ref.~\cite{Kadeer:2005aq} for the single baryon $B_2$ decay. The simplest case is the decay of a spin polarized baryon $B_1\to B_2l^-\bar{\nu}_l$. If the polarization of the final particles is not measured the fully differential angular distribution $d\Gamma\propto{\cal W}=V_{Ph}(q^2)(q^2-m_l^2)\mathrm{Tr}\rho_{B_2}$, where 
 \begin{equation}\label{eq:nSL}
  \mathrm{Tr}\rho_{B_2}\propto \sum_{\mu=0}^3 {C}_{\mu0}\BB_{\mu0}^{B_1B_2}=\sum_{\mu=0}^3 {C}_{\mu0}  \sum_{\kappa=0}^3{\cal R}_{\mu\kappa}^{(4)}( \Omega_2) b_{\kappa0}^{B_1B_2}(q^2,\Omega_l) \ ,
 \end{equation}
with the  baryon $B_1$ spin state in its rest frame described by the polarization vector ${C}_{\mu0}=(1,P_x,P_y,P_z)$. The elements of the decay matrix $b_{\mu0}^{B_1B_2}(q^2,\Omega_l):=b_{\mu0}(q^2,\Omega_l;\boldsymbol\omega_{B_1B_2})$ are given in Eq.~\eqref{eq:bi0}.
For example, if  the initial polarization has only $P_z$ component  the joint angular distribution for the decay process $B_1\to B_2l^-\bar{\nu}_l$ is:
\begin{equation}\label{eq:LLbar_JAD_B}
\begin{aligned}
  \mathcal{W}(\boldsymbol\xi;\boldsymbol\omega)= & V_{Ph}(q^2)(q^2-m_l^2)\left[b_{00}(q^2,\Omega_l;\boldsymbol\omega_{B_1B_2})+P_z b_{30}(q^2,\Omega_l;\boldsymbol\omega_{B_1B_2})\cos\theta_2\right],
  \end{aligned}
\end{equation}
where the vector $\boldsymbol\xi:=(\theta_2,\phi_2,q^2,\Omega_l)$ represents a complete set of the kinematic variables describing an event configuration and the parameter vector $\boldsymbol\omega_{B_1B_2}$ represents the polarization $P_z$, the semileptonic couplings in Eq.~\eqref{eq:g_q2} and the range parameters in Eq.~\eqref{eq:g_q2r}. If the baryon $B_2$ decays weakly as ${B_2}\to{B}_4\pi$ the complete angular distribution is ${\cal W}=V_{Ph}(q^2)(q^2-m_l^2)\mathrm{Tr}\rho_{B_4}$ with
 \begin{equation}\label{eq:nSLp}
  \mathrm{Tr}\rho_{B_4}\propto \sum_{\mu,\nu=0}^3 {C}_{\mu0}\BB_{\mu\nu}^{B_1B_2}a_{\nu0}^{B_2B_4}=\sum_{\mu=0}^3 {C}_{\mu0}  
 \sum_{\kappa,\nu=0}^3{\cal R}_{\mu\kappa}^{(4)}( \Omega_2) b_{\kappa\nu}^{B_1B_2}(q^2,\Omega_l)a_{\nu0}(\theta_4,\phi_4;\alpha_{{B_2}}) \ .
 \end{equation}
The decay matrix $a_{\nu0}(\theta_4,\phi_4;\alpha_{{B_2}})$~\cite{Perotti:2018wxm} describes the non-leptonic decay  ${B_2}\to{B}_4\pi$ and using  the representation from Appendix~\ref{sec:dmexamples} is given as:
\begin{equation}\label{eq:a_coeff_Lbar}
\left[\begin{array}{c}
a_{00}\\
a_{10}\\
a_{20}\\
a_{30}
\end{array}\right]={\cal R}^{(4)}(\{0,\theta_4,\phi_4\})
\left[\begin{array}{c}
1\\
0\\
0\\
\alpha_{{B_2}}
\end{array}\right]=
\left[\begin{array}{l}
1\\
\alpha_{{B_2}}\sin\theta_4\cos\phi_4\\
\alpha_{{B_2}}\sin\theta_4\sin\phi_4\\
\alpha_{{B_2}}\cos\theta_4
\end{array}\right]\ ,
\end{equation}
where $\theta_4$ and $\phi_4$ are the helicity angles of ${B}_4$ in the $\mathbb{R}_2$ frame and $\alpha_{{B_2}}$ is the decay asymmetry parameter. 
The corresponding angular distribution for charge-conjugated decay mode is obtained by the replacements $H_{\lambda_2\ulambda_W}^{B_1}\to H_{\lambda_2\ulambda_W}^{\bar{B}_1}$, $g_{av/w}^{B_1}\to g_{av/w}^{\bar{B}_1}$ and swapping between $(l^-,\bar{\nu}_l)$ and $(l^+,\nu_l)$. Neglecting hadronic CP-violating effects, one has $H_{\lambda_2,\ulambda_W}^{V(\bar{B}_1)}=H_{\lambda_2,\ulambda_W}^{V({B}_1)}$ and 
$H_{\lambda_2,\ulambda_W}^{A(\bar{B}_1)}=-H_{\lambda_2,\ulambda_W}^{A({B}_1)}$ Eq.~\eqref{eq:rel_helampl} meaning that $g_{w}^{\bar{B}_1}=g_{w}^{B_1}$ and $g_{av}^{\bar{B}_1}=-g_{av}^{B_1}$~\cite{Weinberg:1958ut,Frampton:1971sj}.
  
Now we consider a decay of a spin-entangled baryon--antibaryon system $B_1\bar B_1$, where the initial state is given by the spin correlation matrix $C_{\mu\bar{\nu}}^{{B_1}\bar{B}_1}$ defined in Eq.~\eqref{eq:spincor} with $B_1\to B_2 l^-\bar{\nu}_l$. The semileptonic decay is tagged by a common decay of the antibaryon $\bar{B_1}$. For hyperon decay studies, a non-leptonic decay $\bar{B_1}\to\bar{B}_3\bar \pi$ is used.
One obvious advantage of the studies using baryon--antibaryon pairs is that the charge-conjugated decays, 
corresponding to the $\bar B_1\to\bar B_2 l^+{\nu}_l$ and ${B_1}\to{B}_3\pi$ scenario, can be studied simultaneously. A common practice is to implicitly combine events corresponding to the charge-conjugated channels in the analyses to determine the decay properties in the CP-symmetry limit. In such analyses, the quantities that are even (odd) with respect to the parity  operation have the same (opposite sign) values when combining the two cases. At the same time, the CP-symmetry can be tested by comparing values of the separately determined parameters for the baryon and antibaryon decays.
Using as a building block the semileptonic decay matrix one constructs the angular distribution for the case when polarization of baryons $B_2$ and $\bar B_3$ is not measured:
 \begin{equation}\label{eq:LLbar}
  \mathrm{Tr}\rho_{{B_2}\bar B_3}\propto\sum_{\mu,\bar{\nu}=0}^3 C_{\mu\bar{\nu}}^{{B_1}\bar{B}_1}\BB_{\mu0}^{B_1 B_2}a_{\bar{\nu}0}^{\bar{B}_1\bar{B}_3}.
 \end{equation}
The matrix $\BB_{\mu0}^{B_1 B_2}:= \BB_{\mu0}(\theta_2,\phi_2,q^2,\Omega_l;\boldsymbol{\omega}_{B_1B_2})$ describes the semileptonic decay and $a_{\bar\nu0}^{\bar{B}_1\bar{B}_3}:=a_{\bar\nu0}(\theta_3,\phi_3;\bar\alpha_{{B_1}})$~\cite{Perotti:2018wxm} describes the non-leptonic decay  $\bar{B}_1\to\bar{B}_3\pi$, where $\theta_3$ and $\phi_3$ are the helicity angles of $\bar{B}_3$ in the $\bar{B}_1$ rest frame and $\bar\alpha_{{B_1}}$ is the decay asymmetry parameter. 
The joint angular distribution for the process is $\mathcal{W}(\boldsymbol\xi;\boldsymbol\omega)=V_{Ph}(q^2)(q^2-m_l^2) \mathrm{Tr}\rho_{{B_2}\bar B_3}$, where:
\begin{equation}\label{eq:LLbar_JAD}
\begin{aligned}
  \mathrm{Tr}\rho_{{B_2}\bar B_3}= & C_{00}^{{B_1}\bar{B}_1}(\theta_1)b_{00}(\boldsymbol\xi') +
  \sum_{i,j=1}^3 C_{ij}^{{B_1}\bar{B}_1}(\theta_{1})\BB_{i0}(\boldsymbol\xi') a_{j0}(\theta_3,\phi_3;\bar\alpha_{{B_1}})\\
  & +\sum_{i=1}^3 C_{i0}^{{B_1}\bar{B}_1}(\theta_{1}) \BB_{i0}(\boldsymbol\xi')+b_{00}(\boldsymbol\xi')\sum_{j=1}^3 C_{0j}^{{B_1}\bar{B}_1}(\theta_{1})a_{j0}(\theta_3,\phi_3;\bar\alpha_{{B_1}})
  \ 
  \end{aligned}
 \end{equation}
with $C_{\mu\bar\nu}^{{B_1}\bar{B}_1}$ given in Eq.~\eqref{eq:prod} for the annihilation of the unpolarized electron--positron beams. The vectors of the kinematic variables are $\boldsymbol\xi=(\theta_{1},\theta_2,\phi_2,q^2,\Omega_l,\theta_3,\phi_3)$ while  $\boldsymbol\xi'=(\theta_2,\phi_2,q^2,\Omega_l)$. The full vector of parameters is denoted as $\boldsymbol\omega:=(\alpha_{\psi},\Delta\Phi,g_{av}^{B_1},g_w^{B_1},\bar\alpha_{{B_1}})$. 

\section{Sensitivities for SL formfactors parameters}\label{sec:Sens}
\begin{table}
\begin{center}
  \caption{Properties of selected semileptonic decays of the ground-state-octet hyperons. The column labelled $M_1-M_2$ gives the upper range of the $\sqrt{q^2}$ variable. }
    \label{tab:decayproperties-sl}
\begin{ruledtabular}
\begin{tabular}{lcclccl}
    \multirow{2}{*}{Decay}&\multirow{2}{*}{Transition}&\multirow{2}{*}{${\cal B} (\times10^{-4})$}&\multirow{2}{*}{$g_{av}$}&\multirow{2}{*}{$g_w$}& $M_1-M_2$&\multirow{2}{*}{Comment}\\
    &&&&&[MeV]&\\
    \hline
    $\Lambda\to pe^-\bar{\nu}_e$&$V_{us}$&$8.32(14)$&$0.718(15)$&$1.066$& $177$& \cite{Workman:2022ynf,Cabibbo:2003cu}\\
    $\Sigma^+\to\Lambda e^+{\nu}_e$\footnotemark[1]&$V_{ud}$&$0.20(05)$&$0.01(10)$&$\phantom{-}2.4(17)$&$\phantom{0}74$& \cite{Workman:2022ynf}\\
    $\Xi^-\to \Lambda e^-\bar{\nu}_e$&$V_{us}$&$5.63(31)$&$0.25(5)$&$0.085$&$206$& \cite{Bourquin:1983de,Cabibbo:2003cu}\\
    \hline
    $\Xi^-\to\Sigma^0 e^-\bar{\nu}_e$&$V_{us}$&$0.87(17)$&$1.25(15)$&$2.609$&$129$& \cite{Bourquin:1983de,Cabibbo:2003cu}\\
    $\Xi^0\to\Sigma^+ e^-\bar{\nu}_e$&$V_{us}$&$2.52(8)$&$1.22(5)$&$2.0(9)$&$125$& \cite{Workman:2022ynf}\\
\end{tabular}
\end{ruledtabular}
\end{center}
\vspace{-0.5cm}
\noindent\begin{minipage}[t]{\textwidth}
\footnotetext[1]{Since for $\Sigma^+$ $F_1=0$, the coupling constants $g_{av}$ and $g_w$ are defined as $F_1^V/F_1^A$ and $F_2^V/F_1^A$, respectively.}
\end{minipage}
\end{table}
Here we present estimates for the statistical uncertainties of the parameters describing formfactors of selected semileptonic hyperon decays. The derived angular distributions are used to construct the normalized multidimensional probability density function for an event configuration. They are functions of $q^2$ and the helicity angles, and depend on the formfactor parameters such as $g_{av}$ and $g_w$ \eqref{eq:g_q2}.
The parameters can be determined in an experiment using maximum likelihood (ML) method, which guarantees  consistency and efficiency properties. We provide uncertainties of the parameters in the large number of events limit and assuming the detection efficiency does not depend on the kinematic variables as described in Refs.~\cite{Adlarson:2019jtw,Salone:2022lpt}. Since the ML estimators are asymptotically normal, the  product of their standard deviations, $\sigma$, and $\sqrt{N}$, where $N$ is the  number of the observed events, does not depend on $N$. The uncertainties are obtained by calculating elements of the Fisher information matrix that is inverted to obtain the covariance matrix for the parameters.

\begin{figure}
\centering
   \includegraphics[width=0.80\columnwidth]{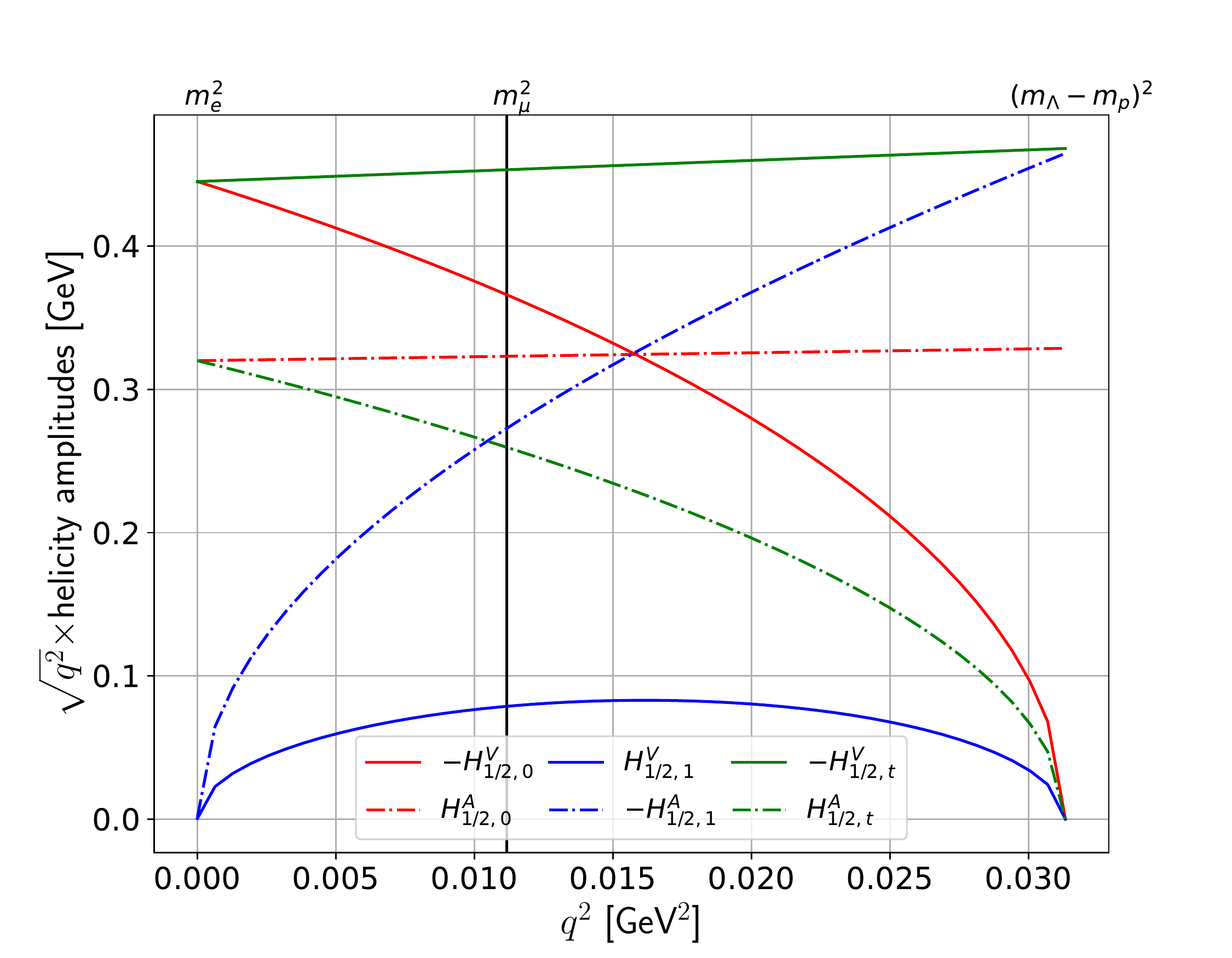}
\put(-380,240){\Large(a)} 

\includegraphics[width=0.80\columnwidth]{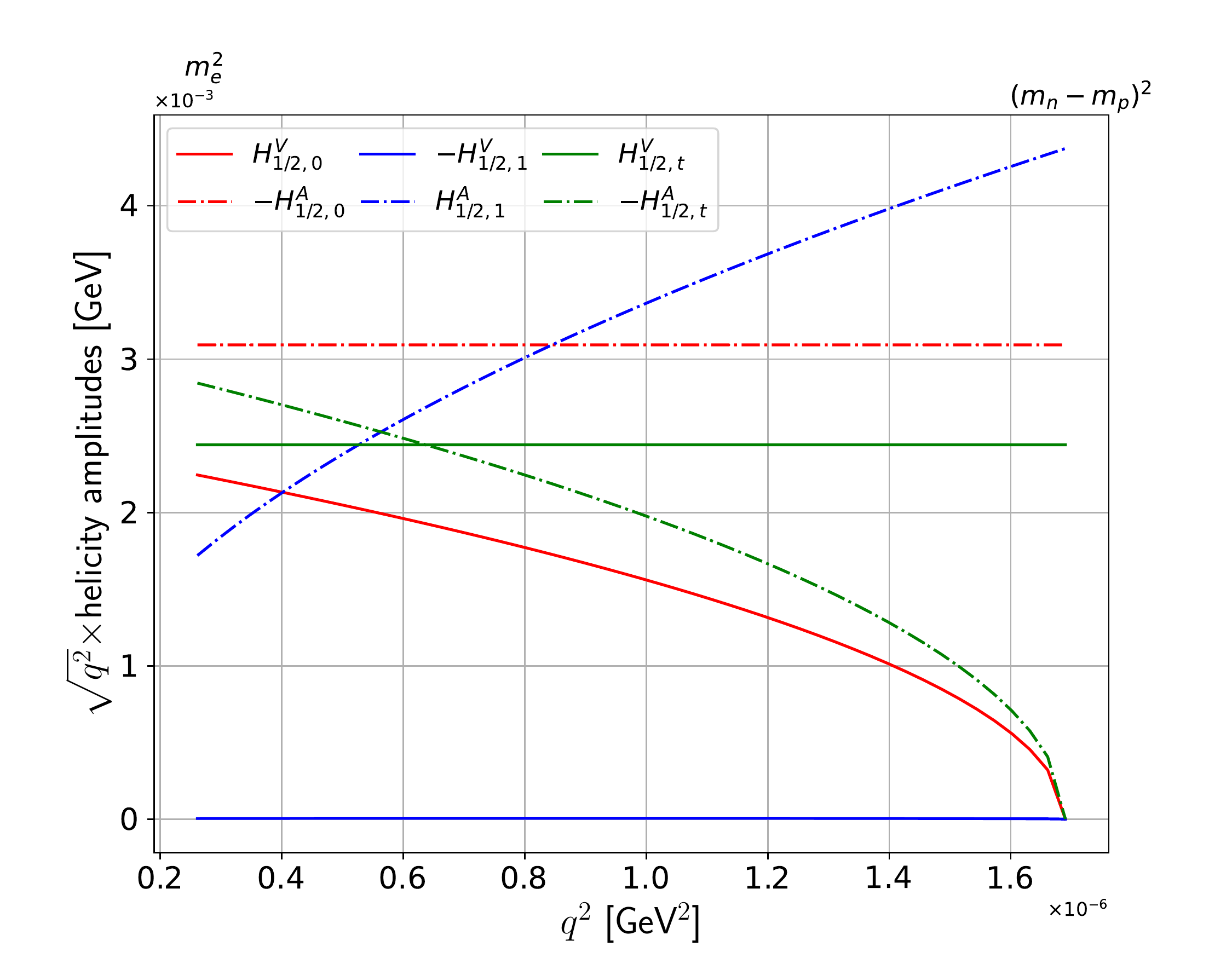}
\put(-380,240){\Large(b)} 
\caption{The $q^2$-dependence of the six  helicity amplitudes for (a) $\Lambda\to pl^-\bar{\nu}_l$ and (b) $n\to pe^-\bar{\nu}_e$ decays. For the $\Lambda$ decay the kinematic range for the $\mu$-mode  is to the right of the vertical line $q^2=m^2_{\mu}$. 
}
 \label{fig:helampl}
\end{figure}

We consider  the semileptonic decays of hyperons listed in Table~\ref{tab:decayproperties-sl}. 
We neglect formfactors $F_3^V$ and $F_2^A$ which vainsh in the limit of of the SU(3) flavor symmetry~\cite{Marshak69}.
Equation~\eqref{eq:helamp_semil} allows one to estimate the relative contribution of different formfactors to the angular distributions. Based on the $g_{av}$ and $g_w$ values from Table~\ref{tab:decayproperties-sl} the $q^2$ dependence of the six helicity amplitudes for the $\Lambda$ semileptonic decays is shown in Fig.~\ref{fig:helampl}(a). To allow a better comparison the amplitudes are multiplied by $\sqrt{q^2}$. Close to the lower boundary, $q^2=m^2_e$, the longitudinal and scalar helicity amplitudes dominate, with $H^{V(A)}_{\frac{1}{2}0}\approx H^{V(A)}_{\frac{1}{2}t}$. Close to the upper boundary at the zero recoil point, $q^2=(M_1-M_2)^2$, the contributions $H^V_{\frac{1}{2}t}$ and $H^A_{\frac{1}{2}1}=-\sqrt{2}H^A_{\frac{1}{2}0}$ are dominant with $H^V_{\frac{1}{2}t}=-H^A_{\frac{1}{2}0}/g_{av}$. 
We do not consider the decay $\Sigma^-\to ne^-\bar{\nu}_e$ since the final state includes two neutral particles, neutron and neutrino, making it impossible to fully reconstruct the events. In addition, no measurements exist for the production parameters in  the $e^+e^-\to\Sigma^-\bar{\Sigma}^+$ process. 

The first case is the decay $\Lambda\to pe^-\bar{\nu}_e$ studied in the exclusive process $e^+e^-\to J/\psi\to\Lambda\bar{\Lambda}$, where $\bar{\Lambda}\to\bar{p}\pi^+$ is used for tagging. The angular distribution is given by Eq.~\eqref{eq:LLbar} where the parameters of the production process $e^+e^-\to J/\psi\to\Lambda\bar{\Lambda}$ needed to define the spin-correlation--polarization matrix $C_{\mu\nu}$ are given in Table~\ref{tab:Prod}. The properties of $\bar{\Lambda}\to\bar{p}\pi^+$ decay and the charge conjugated process that is used to tag the SL decay are given in Table~\ref{tab:decayproperties}. We assume the production parameters and the decay parameters of the non-leptonic decays used for the tagging to be well known and fixed. Since the coupling $g_{av3}$ is multiplied by $m_e$ in the transition amplitude~\cite{Goldberger:1958vp}, we set it to zero because it cannot be determined from experiment with a reasonable uncertainty. In addition the parameters $r_{v,w}$ and $r_{av}$ defined in Eq.~\eqref{eq:g_q2r} are fixed to the values deduced from the ansatz for the $s\to u$ transition of Refs.~\cite{Gaillard:1984ny,Cabibbo:2003cu} and listed in Table~\ref{tab:sens}.
\begin{table}
  \caption{Properties of the $e^+e^-\to J/\psi\to B_1\overline B_1$ decays to the pairs of ground-state octet hyperons. \label{tab:Prod}}
\begin{ruledtabular}
\begin{tabular}{lllll}
  &${\cal B}(\times 10^{-4})$& $\alpha_\psi$&$\Delta\Phi$ [rad]&Comment\\ \hline
$\Lambda\overline{\Lambda}$&$19.43(33)$&$\phantom{-}0.475(4)$&$\phantom{-}0.752(8)$&\cite{BESIII:2018cnd,BESIII:2017kqw}\\
$\Sigma^+\overline\Sigma\vphantom{X}^-$&$15.0(24)$&$-0.508(7)$&$-0.270(15)$ &\cite{BES:2008hwe, BESIII:2020fqg}\\
$\Xi^-\overline\Xi\vphantom{X}^+$&$\phantom{0}9.7(8)$&$\phantom{-}0.586(16)$&$\phantom{-}1.213(49)$&\cite{Workman:2022ynf,BESIII:2021ypr}\\
{{$\Xi^0\overline\Xi\vphantom{X}^0$}}
& $11.65(4)$&$\phantom{-}0.514(16)$&$\phantom{-}1.168(26)$& \cite{BESIII:2016nix,BESIII:2023drj}\\
  \end{tabular}
\end{ruledtabular}
\end{table}
\begin{table}
\begin{center}
  \caption{Properties of the main decays of the ground-state octet hyperons that can be used to tag the SL decays.  
  The decay asymmetry $\bar\alpha_D$ for the charge conjugated  decay modes in the CP-symmetry conservation limit is  $\bar\alpha_D=-\alpha_{D}$.  }
    \label{tab:decayproperties}
\begin{ruledtabular}
\begin{tabular}{lccc}
    \hfill$D$&${\cal B}(\%)$&${\alpha_D}$&Comment\\
    \hline
    $\Lambda\to p\pi^-$&\phantom{0}$64$&$\phantom{-}{0.755(3)}$& \cite{BESIII:2021ypr,BESIII:2022qax}\\
    $\Sigma^+\to p\pi^0$&\phantom{0}${52}$&${-0.994(4)}$ & \cite{BESIII:2020fqg}\\
    $\Xi^-\to \Lambda\pi^-$&$100$&${-0.379(4)}$&\cite{Workman:2022ynf,BESIII:2021ypr}\\
    $\Xi^0\to \Lambda\pi^0$&$96$&${-0.375(3)}$&\cite{Workman:2022ynf,BESIII:2023drj}   
\end{tabular}
\end{ruledtabular}
\end{center}
\end{table}
\begin{table}
  \caption{Statistical uncertainties for the $g_{av}$ and $g_w$ couplings for some semileptonic decays reconstructed using double-tag method Eq.~\eqref{eq:LLbar}.  \label{tab:sens}}
\begin{ruledtabular}
\begin{tabular}{l|ll|ll}
    Decay&$\sigma(g_{av}){\sqrt{N}}$&$\sigma(g_w){\sqrt{N}}$&$r_{v,w}$ [GeV$^{-2}$] & $r_a$ [GeV$^{-2}$]\\
    \hline
    $\Lambda\to pe^-\bar{\nu}_e$&$1.8$&$12$&\multirow{5}{*}{$1.94$} & \multirow{5}{*}{$1.28$}\\
    $\Xi^-\to[\Lambda\to pe^-\bar{\nu}_e]\pi^-$&$1.8$&$12$&&\\
    $\Xi^-\to[\Lambda\to p\pi^-]e^-\bar{\nu}_e$&$0.6$&$\phantom{0}9$&&\\
     {$\Xi^-\to[\Sigma^0\to[\Lambda\to p\pi^-]\gamma] e^-\bar{\nu}_e$}&{$5.0$}&{$29$}& &\\
    {$\Xi^0\to[\Sigma^+\to p\pi^0] e^-\bar{\nu}_e$}&{$4.0$}&{$28$}&&\\
    \hline
    $\Sigma^+\to[\Lambda\to p\pi^-] e^+\nu_e$&$0.5$&$19$&$2.83$&$1.71$\\
  \end{tabular}\\
\end{ruledtabular}
\end{table}
The statistical uncertainties $\sigma(g_{av})$ and $\sigma(g_w)$ for the coupling constants $g_{av}$ and $g_w$, respectively, are given in the first row of Table~\ref{tab:sens}. The main feature is that the uncertainty for the $g_{av}$ coupling is nearly one order of magnitude less than for $g_w$ since the latter is suppressed by the $q^2/M^2_1<(M_1-M_2)^2/M_1^2\approx0.025$ factor~\eqref{eq:helamp_semil}. The second row corresponds to an independent method to study $\Lambda\to pe^-\bar{\nu}_e$ using the $e^+e^-\to J/\psi\to\Xi^-\bar{\Xi}^+$ process with  the  $\Xi^-\to[\Lambda\to pe^-\bar{\nu}_e]\pi^-$ sequence and  $\bar\Xi^+\to[\bar\Lambda\to \bar p\pi^+]\pi^-$ for the tagging. The modular expression for the angular distribution of such process reads
 \begin{equation}\label{eq:XXbarL}
  \mathrm{Tr}\rho_{{p}\bar p}\propto\sum_{\mu,\bar{\nu}=0}^3 C_{\mu\bar{\nu}}^{{\Xi}\bar{\Xi}}\sum_{\mu'=0}^3a_{{\mu}\mu'}^{\Xi\Lambda}\BB_{\mu'0}^{\Lambda p} \sum_{\bar{\nu}'=0}^3a_{\bar{\nu}\bar\nu'}^{\bar{\Xi}\bar{\Lambda}}a_{\bar\nu'0}^{\bar\Lambda\bar p} \ .
 \end{equation}
The polarization of the $\Lambda$ originating from the non-leptonic weak decay $\Xi^-\to\Lambda\pi^-$, is $\sim40\%$, to be compared to the root-mean-squared value of the $\Lambda$ polarization in $e^+e^-\to J/\psi\to\Lambda\bar{\Lambda}$ of $11\%$~\cite{Salone:2022lpt}. However, the uncertainties of the weak couplings are the same for both methods. To further investigate dependence on the initial polarization of $\Lambda$  we set $\Delta\Phi=0$ to have the zero polarization, while to obtain maximally polarized $\Lambda$ we include the longitudinal polarization of the electron beam and use the production matrix $C_{\mu\bar\nu}$ from Ref.~\cite{Salone:2022lpt}. The impact of the spin correlations for the uncertainties can be studied by comparing the results using the  angular distributions \eqref{eq:LLbar} or  \eqref{eq:XXbarL} with full production matrices $C_{\mu\bar\nu}$ to the ones where all elements except $C_{\mu0}$ are set to zero. This arrangement assures that the spin correlation terms are excluded.  In all these tests the  uncertainties of $\sigma(g_{av})$ and $\sigma(g_w)$ remain unchanged, meaning that the polarization and the spin correlations of the mother hyperon in the decay play almost no role for the measurements of properties of the semileptonic decays to baryons whose polarization is not measured.

\begin{figure}
\centering
\includegraphics[width=0.8\columnwidth]{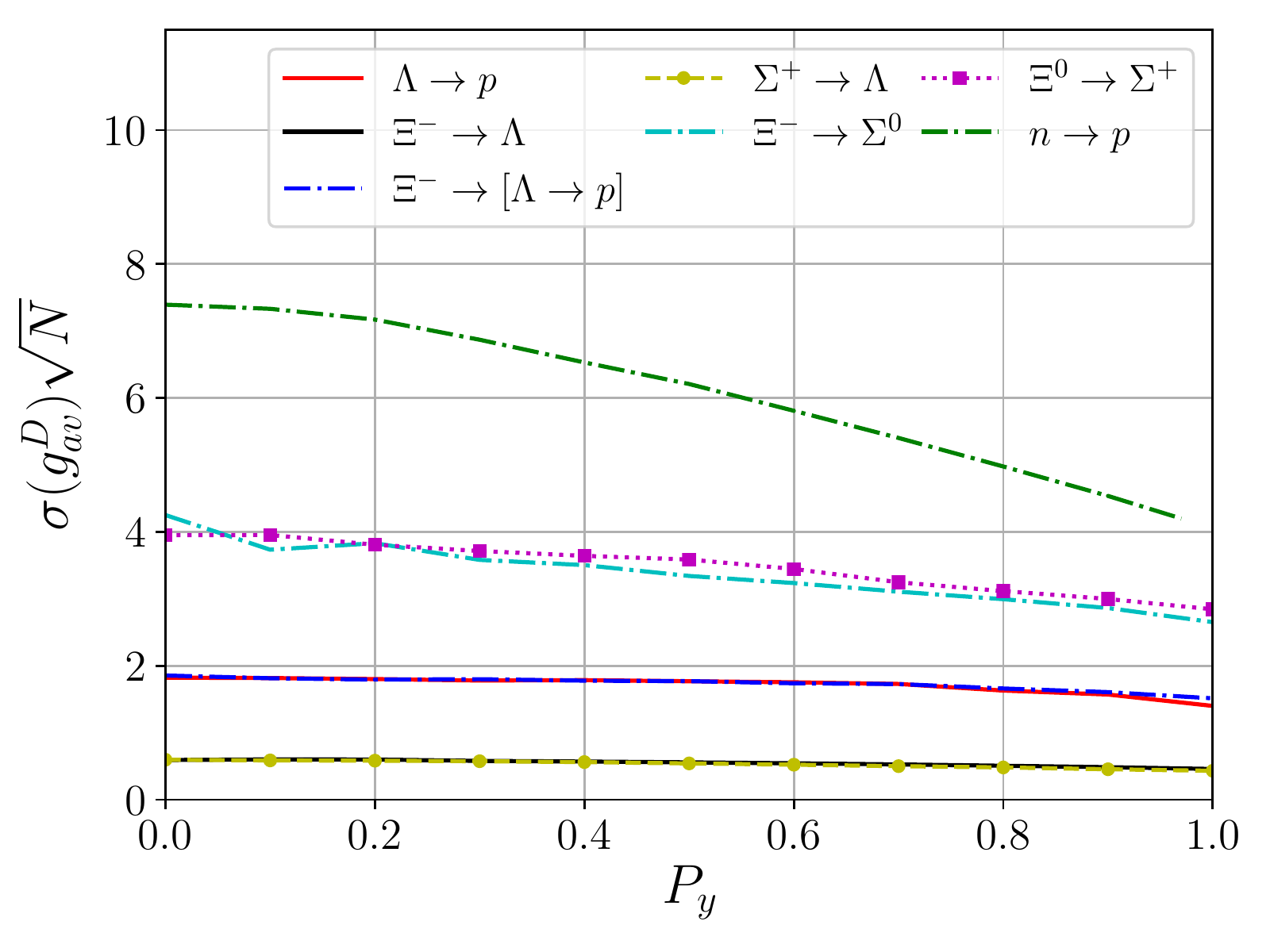}
\put(-380,240){\Large(a)} 

\includegraphics[width=0.8\columnwidth]{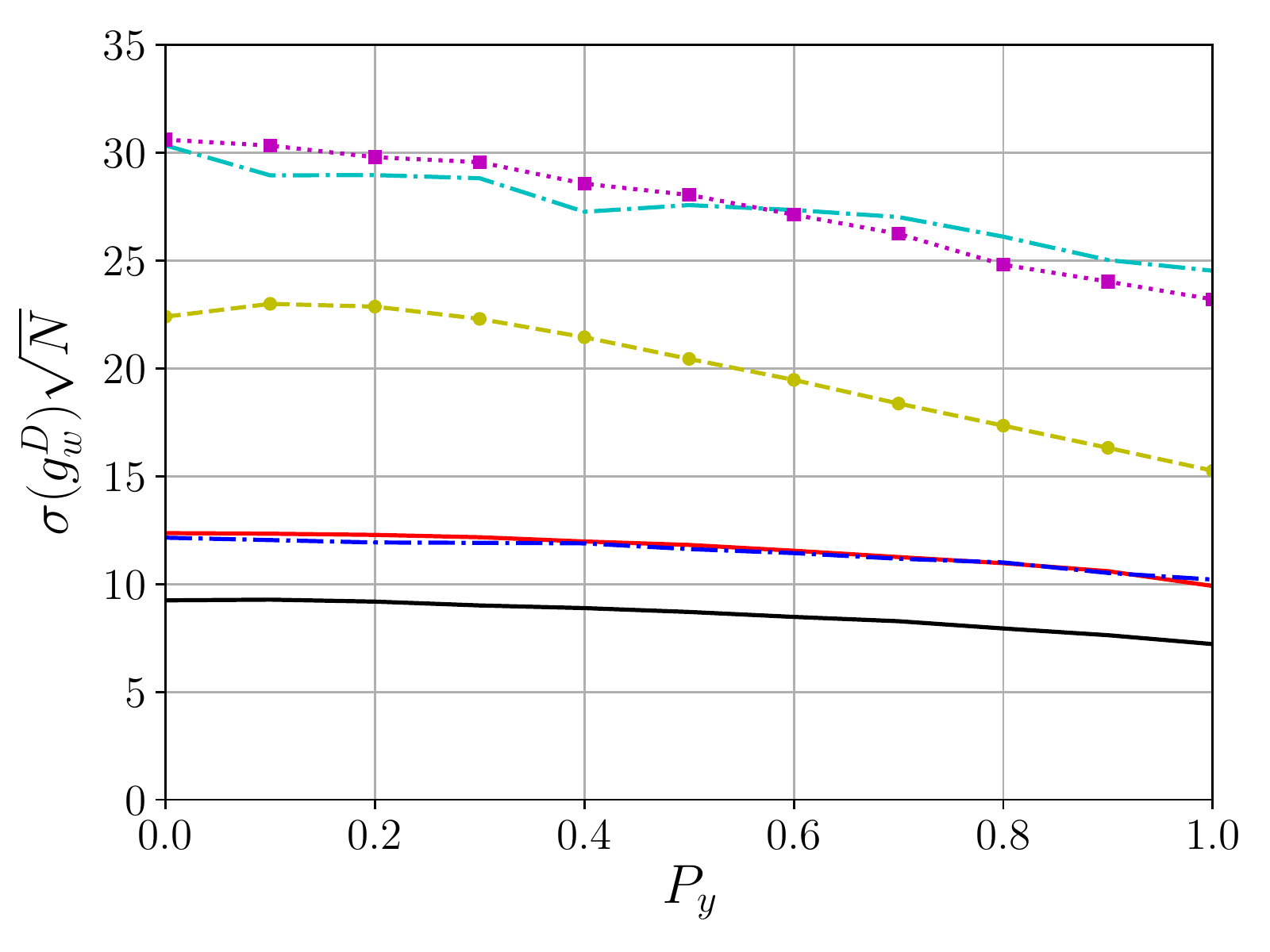}
\put(-380,240){\Large(b)} 
\caption{Statistical uncertainties (a) $\sigma(g_{av})\sqrt{N}$ and (b) $\sigma(g_w)\sqrt{N}$ for semileptonic decays decays as a function of the initial baryon polarization. Note that there is no estimate of $\sigma(g_w)$ for $n\to p e^-\nu_e$ in panel (b) as explained in the text.}
 \label{fig:gavn}
\end{figure}

The entries from the third row and below in Table~\ref{tab:sens} correspond to the decays where the polarization of the daughter baryon is measured and the angular distributions include the complete $\BB_{\mu\mu'}$ matrices. For example the angular distribution for $\Xi^-\to\Lambda e^-\bar{\nu}_e$ measurement in  $e^+e^-\to J/\psi\to\Xi^-\bar{\Xi}^+$ is
\begin{equation}\label{eq:XXbar}
  \mathrm{Tr}\rho_{{p}\bar p}\propto\sum_{\mu,\bar{\nu}=0}^3 C_{\mu\bar{\nu}}^{{\Xi}\bar{\Xi}}\sum_{\mu'=0}^3\BB_{{\mu}\mu'}^{\Xi\Lambda}a_{\mu'0}^{\Lambda p} \sum_{\bar{\nu}'=0}^3a_{\bar{\nu}\bar\nu'}^{\bar{\Xi}\bar{\Lambda}}a_{\bar\nu'0}^{\bar\Lambda\bar p} \ .
 \end{equation}
Since the uncertainties  depend on the values of the weak couplings it is difficult to compare the results for different decays in Table~\ref{tab:sens}. 
By repeating the studies with variation of $\Delta\Phi$ and the electron beam polarization some impact is seen for the uncertainties, specially for the $g_{w}$ parameter in $\Sigma^+\to\Lambda e^+\nu_e$.
In addition  we study the uncertainties for single spin polarized baryon decays with the angular distributions given by Eq.~\eqref{eq:nSL}.  The baryon $B_1$  polarization vector is set to ${C}_{\mu0}=(1,0,P_y,0)$. The results for $\sigma(g_{av})$ and $\sigma(g_w)$ are shown in Fig.~\ref{fig:gavn}. The uncertainty for large $P_y$ decreases typically by 20\% comparing to the unpolarized case.

Our formalism applies also to the $n\to pe^-\bar{\nu}_e$ decay and  it should be equivalent to the approach from Ref.~\cite{Groote:2019rmj} for the single neutron decay. However, we can also describe decay correlations for a spin entangled neutron--neutron pair. As an example we take $nn$ spin singlet state given by the spin correlation matrix $C_{\mu\nu}=\text{diag}(1,-1,1,1)$. The coupling constants $g_{av}$ and $g_{w}$ are  $1.2754(13)$ and $1.853$, respectively \cite{Workman:2022ynf,Groote:2019rmj}. The $q^2$ dependence of the formfactors is neglected due to the tiny range, $m_e<\sqrt{q^2}<M_n-M_p$, of the variable. The corresponding helicity amplitudes for the neutron beta decay are shown in Fig.~\ref{fig:helampl}(b). The resulting  uncertainty of the $g_{av}$ measurement in the double beta decay of the singlet pair is $\sigma(g_{av})\sqrt{N}=4.3$. It should be compared to the uncertainties in the measurements with single neutrons that are shown in Fig.~\ref{fig:gavn}(a) as a function of the neutron polarization. For unpolarized neutron $\sigma(g_{av})\sqrt{N}=7.4$ and it decreases to 4.1 when the polarization is equal one. The flip-contribution to the helicity amplitudes~\eqref{eq:flip} of about 8\% was neglected in the estimates. The $g_w$ coupling cannot be determined since its contribution to the helicity amplitude $H^V_{\frac{1}{2}0}$ is suppressed by a factor $q^2/M_n^2$. Moreover, the second amplitude that includes $g_w$,  $H^V_{\frac{1}{2}1}$, is suppressed by $\sqrt{q^2}$ and as  seen in Fig.~\ref{fig:helampl}(b) it is  consistent with zero.

\section{Conclusions}

We have constructed a modular description of the differential distributions for baryon semileptonic decays where the baryons are originating from entangled baryon--antibaryon pairs produced in the electron--positron annihilations or in charmonia decays. The formalism allows to extract the weak formfactors using complete information available in such experiments.
The lepton mass effects as well as polarization effects of the decaying parent hyperon are included in the formalism. The presented modular expressions are applicable to various sequential processes like  $B_1\to B_2(\to B_3+\pi)+l+\nu_l$ that involve a semileptonic decay. 
Two conventions for defining transversal directions of the helicity frames were considered. The daughter baryon spin-density matrix in a semileptonic decay takes the simplest form when expressed using the angles in the decay plane. The two representations are equivalent, provided that one uses the matching set of rotations to define the helicity angles.

We have not included radiative corrections in our estimates but they have to be considered in the experimental analyses. Over the years, the radiative corrections to hadronic $\beta$-decays have been extensively studied~\cite{RevModPhys.85.263} and the specific applications to the hyperon semileptonic decays are discussed in Ref.~\cite{Garcia:1985xz}. The state-of-the-art in experimental analyses is to use \textsc{Photos} program~\cite{Barberio:1993qi} that is based on leading-logarithmic (collinear) approximation. The procedure is applied to all final particles but the electron(positron) tracks are most affected.

The BESIII experiment has collected $10^{10}$ $J/\psi$~\cite{BESIII:2021cxx} meaning that for semileptonic decays data samples of less than $10^{4}$ events are available. Therefore a rough estimate of the achievable uncertainties with this data set is given by dividing the values in Table~\ref{tab:sens} by 100. The $J/\psi$ decays into a hyperon--antihyperon pair can provide a clean setting with low systematic uncertainties for the CP-symmetry conservation tests in semileptonic decays since the decays of the charge conjugated modes can be done simultaneously.

A similar modular approach with decay matrices might be useful for studies of radiative and Dalitz decays. As a cross-check and illustration in Appendix~\ref{sec:dmexamples} we provide formulas for the Dalitz transition $B_1\to B_2 l^+l^-$ between baryons with spin 1/2  as well as decay matrix for a weak radiative decay with real photon $B_1\to B_2\gamma$.

For the studies of semileptonic decays of heavy baryons induced by the quark transitions $c\to s+l^++\nu_l$ or $b\to c+l^-+\bar{\nu}_l$ the previously available formalism~\cite{Kadeer:2005aq} is likely sufficient if only beams of polarized baryons are used. This might change in near future with BESIII and Belle II experiments where entangled charmed baryon--antibaryon pairs will be  available. One difference would be a measurement of the polarization for the tagging reactions which probably has to use three-body hadronic weak decays. However, even for the case of single baryon decays our approach provides an easy and flexible way to implement different decay sequences in the event generators that propagate spin information of the decaying baryons.

\begin{acknowledgments}
We thank our colleagues Stefan Leupold and Patrik Adlarson for useful discussion and for motivation of this study. This work was supported in part by National Natural Science Foundation of China (NSFC) under Contract No. 11935018, the CAS President’s International Fellowship Initiative (PIFI) Grant No. 2021PM0014, {Polish National Science Centre through the Grant No. 2019/35/O/ST2/02907} and Swedish Research Council through the Grant No. 2021-04567.  In addition, a support from the European Union's Horizon 2020 research and innovation programme under grant agreement STRONG-2020 -- No 824093 and the Munich Institute for Astro-, Particle and BioPhysics (MIAPbP) which is funded by the Deutsche Forschungsgemeinschaft (DFG, German Research Foundation) under Germany's Excellence Strategy - EXC-2094 - 390783311 is acknowledged. 
\end{acknowledgments}

\appendix
\section{Conventions for Wigner functions and Pauli matrices}\label{app:pauli}
Conventions for Pauli matrices: rows $m_1=1/2,-1/2$ are numbered from top to bottom and columns $m_2=1/2,-1/2$ are numbered from left to right:
\begin{equation}
\sigma_0^{m_1,m_2}=\left(
\begin{array}{rr}
 1 & 0 \\
 0 & 1 \\
\end{array}
\right) ,\
\sigma_1^{m_1,m_2}=\left(
\begin{array}{rr}
 0 & 1 \\
 1 & 0 \\
\end{array}
\right) ,\
\sigma_2^{m_1,m_2}=\left(
\begin{array}{rr}
 0 & -i \\
 i & 0 \\
\end{array}
\right) ,\
\sigma_3^{m_1,m_2}=\left(
\begin{array}{rr}
 1 & 0 \\
 0 & -1 \\
\end{array}
\right)\ .
\end{equation}
The corresponding Wigner functions $\mathcal{D}^J_{m_1,m_2}(0,\theta,0)=d^J_{m_1,m_2}(\theta)$  are for $J=1/2$
\begin{equation}
{\cal D}^{1/2}_{m_1,m_2}(0,\theta,0)=    \left(
\begin{array}{rr}
 \cos {\theta }/{2} & -\sin {\theta }/{2} \\
 \sin {\theta }/{2} & \cos {\theta }/{2} \\
\end{array}
\right)\ , 
\end{equation}
with the columns and rows expressed using the same convention. 
For $J=1$ the functions are 
\begin{equation}
{\cal D}^{1}_{m_1,m_2}(0,\theta,0)=    \left(
\begin{array}{rrr}
\frac{1}{2}(1+\cos\theta) & \frac{1}{\sqrt{2}}\sin\theta & \frac{1}{2}(1-\cos\theta) \\
-\frac{1}{\sqrt{2}}\sin\theta & \cos\theta & \frac{1}{\sqrt{2}}\sin\theta \\
\frac{1}{2}(1-\cos\theta) & -\frac{1}{\sqrt{2}}\sin\theta & \frac{1}{2}(1+\cos\theta)\\
\end{array}
\right) \ ,
\end{equation}
where the rows ($m_1$) and columns ($m_2$) are labeled in the order $(-1, 0, 1)$ from left to right and top to bottom, respectively. This convention matches the complete Wigner ${\cal D}$ functions given as $\mathcal{D}^J_{m_1,m_2}(\phi,\theta,\chi)=\exp(-im_1\phi){\cal D}^{J}_{m_1,m_2}(0,\theta,0)\exp(-im_2\chi)$. The Pauli matrices are related to the 3D rotation matrices $R_{jk}(\Omega)$ for $\Omega\equiv\{\phi,\theta,\chi\}$ in the following way :
\begin{equation}
 \begin{split}
R_{jk}(\Omega)&= \frac12\sum_{\kappa,\kappa'}\sum_{\zeta,\zeta'}~
(\sigma_j)^{\kappa,\kappa'}(\sigma_k)^{\zeta',\zeta}{\cal D}^{1/2 *}_{\kappa,\zeta}(\Omega)
{\cal D}^{1/2}_{\kappa',\zeta'}(\Omega)\\
&=\left(
\begin{array}{ccc}
 \cos \theta \cos \chi  \cos \phi -\sin \chi  \sin \phi  & -\cos \theta  \sin \chi  \cos \phi -\cos \chi  \sin \phi  & \sin \theta  \cos \phi  \\
  \cos \theta  \cos \chi  \sin \phi +\sin \chi  \cos \phi  & \cos \chi  \cos \phi -\cos \theta  \sin \chi  \sin \phi  & \sin \theta \sin \phi  \\
  -\sin \theta  \cos \chi  & \sin \theta \sin \chi  & \cos \theta  \\
\end{array}
\right) ,
\end{split}   
\end{equation}
where the columns ($k$) and rows ($j$) are labeled $k,j=1,2,3$ ($x,y,z$) from left to right and from top to bottom, respectively.
\section{Derivation of the decay matrix decomposition}\label{sec:DerivRBb}
Starting from the amplitude representation in Eq.~\eqref{eq:DHseparate} we derive expression Eq.~\eqref{eq:BRb}.
Multiplying the amplitude in Eq.~\eqref{eq:DHseparate}
by its conjugate to obtain spin-density matrix and by inserting basis Pauli matrices for the mother and the daughter baryon:
\begin{equation}
\begin{split}
\BB^D_{\mu\nu}&=\frac{1}{8\pi^2} \sum_{\lambda,\lambda'}\sum_{\kappa,\kappa'}\sum_{\zeta,\zeta'}\HH_{\zeta,\lambda}\HH^*_{\zeta',\lambda'}~
(\sigma_\mu)^{\kappa,\kappa'}(\sigma_\nu)^{\lambda',\lambda}{\cal D}^{1/2 *}_{\kappa,\zeta}(\Omega)
{\cal D}^{1/2}_{\kappa',\zeta'}(\Omega)\\
&=\frac{1}{8\pi^2} \sum_{\zeta,\zeta'}
\underbrace{\left[\sum_{\lambda,\lambda'}(\sigma_\nu)^{\lambda',\lambda}\HH_{\zeta,\lambda}\HH^*_{\zeta',\lambda'}\right]}_{I_\nu^{\zeta,\zeta'}}\underbrace{\left[\sum_{\kappa,\kappa'}(\sigma_\mu)^{\kappa,\kappa'}{\cal D}^{1/2 *}_{\kappa,\zeta}(\Omega){\cal D}^{1/2}_{\kappa',\zeta'}(\Omega)\right]}_{R_\mu^{\zeta,\zeta'}(\Omega)}\\
&=\frac{1}{8\pi^2} \sum_{\zeta,\zeta'}R_\mu^{\zeta,\zeta'}(\Omega)I_\nu^{\zeta,\zeta'}\ ,
\end{split}\label{eq:h2sig1}
\end{equation} 
where the running indices in all sums  $\kappa,\kappa',\lambda',\lambda$ and $\zeta,\zeta'$ are $-1/2$ and $+1/2$. Despite  $\BB^D_{\mu\nu}$ being a real-value matrix,  the matrices $I_\nu^{\zeta,\zeta'}$ and $R_\mu^{\zeta,\zeta'}(\Omega)$ are not real-valued. 
We would like to rewrite Eq.~\eqref{eq:h2sig1} as a product of a 4D rotation matrix and a $4\times4$ matrix $b_{\rho\nu}$ in the form given in Eq.~\eqref{eq:BRb}:
\begin{equation*}
\begin{split}
\BB^D_{\mu\nu}=\frac{1}{8\pi^2} \sum_{\rho=0}^3{\cal R}_{\mu\rho}^{(4)}(\Omega)b_{\rho\nu} \ .
\end{split}
\end{equation*} 
In order to derive form of ${\cal R}_{\mu\rho}^{(4)}(\Omega)$ we set the matrix $b_{\mu\nu}$ to the identity $
4\times4$ matrix. This can be achieved by setting $\HH_{\zeta,\lambda}=\delta_{\zeta\lambda}$ since
\begin{equation}
\begin{split}
\sum_{\zeta,\zeta'}
{\sum_{\lambda,\lambda'}(\sigma_\nu)^{\lambda',\lambda}\delta_{\zeta,\lambda}\delta_{\zeta',\lambda'}}{
(\sigma_\mu)^{\zeta,\zeta'}}=2\delta_{\nu\mu}\ .
\end{split}
\end{equation} 
Such replacement in Eq.~\eqref{eq:h2sig1} gives:
\begin{equation}
\begin{split}
{\cal R}_{\mu\nu}^{(4)}(\Omega)&= \frac12\sum_{\lambda,\lambda'}\sum_{\kappa,\kappa'}\sum_{\zeta,\zeta'}\delta_{\zeta,\lambda}\delta_{\zeta',\lambda'}~
(\sigma_\mu)^{\kappa,\kappa'}(\sigma_\nu)^{\lambda',\lambda}{\cal D}^{1/2 *}_{\kappa,\zeta}(\Omega)
{\cal D}^{1/2}_{\kappa',\zeta'}(\Omega)\\
&= \frac12\sum_{\kappa,\kappa'}\sum_{\zeta,\zeta'}~
(\sigma_\mu)^{\kappa,\kappa'}(\sigma_\nu)^{\zeta',\zeta}{\cal D}^{1/2 *}_{\kappa,\zeta}(\Omega)
{\cal D}^{1/2}_{\kappa',\zeta'}(\Omega)\ .
\end{split}
\end{equation} 
By evaluating the above expression one gets the explicit form for ${\cal R}_{\mu\nu}^{(4)}(\Omega)$:
\begin{equation}
   \left(
\begin{array}{cccc}
 1 & 0 & 0 & 0 \\
 0 & \cos \theta  \cos \chi  \cos \phi -\sin \chi  \sin \phi  & -\cos \theta  \sin \chi \cos \phi -\cos \chi  \sin \phi  & \sin \theta \cos \phi  \\
 0 & \cos \theta  \cos \chi  \sin \phi +\sin \chi  \cos \phi  & \cos \chi  \cos \phi -\cos \theta  \sin \chi  \sin \phi  & \sin \theta \sin \phi  \\
 0 & -\sin \theta  \cos \chi  & \sin \theta \sin \chi  & \cos \theta  \\
\end{array}
\right)\ ,
\end{equation}
which is the 4D rotation where the spatial part ${\cal R}_{jk}(\Omega)$ corresponds to the product of the following three axial rotations:
\begin{equation}
\begin{split}
    {\cal R}_{jk}(\Omega)&=R_z(\phi)R_y(\theta)R_z(\chi)\\
    &=\left(
\begin{array}{ccc}
 \cos \phi  & -\sin \phi  & 0 \\
 \sin \phi  & \cos \phi  & 0 \\
 0 & 0 & 1 \\
\end{array}
\right)\left(
\begin{array}{ccc}
 \cos \theta  & 0 & \sin \theta  \\
 0 & 1 & 0 \\
 -\sin \theta  & 0 & \cos \theta  \\
\end{array}
\right)\left(
\begin{array}{ccc}
 \cos \chi  & -\sin \chi  & 0 \\
 \sin \chi  & \cos \chi  & 0 \\
 0 & 0 & 1 \\
\end{array}
\right)\ .
\end{split}
\end{equation}

The expression for $b_{\mu\nu}$ can be deduced by setting ${\cal R}_{\mu\rho}^{(4)}(\Omega)$ to the $4\times4$ identity matrix i.e. by setting $\Omega=\{0,0,0\}$:
\begin{equation}\label{eq:b_3body}
\begin{split}
b_{\mu\nu}&=\BB^D_{\mu\nu}(\Omega\equiv0)= \sum_{\zeta,\zeta'}
{\left[\sum_{\lambda,\lambda'}(\sigma_\nu)^{\lambda',\lambda}\HH_{\zeta,\lambda}\HH^*_{\zeta',\lambda'}\right]}{
\left[\sum_{\kappa,\kappa'}(\sigma_\mu)^{\kappa,\kappa'}\delta_{\kappa,\zeta}
\delta_{\kappa',\zeta'}\right]}\\
&= \sum_{\zeta,\zeta'}
{\sum_{\lambda,\lambda'}(\sigma_\nu)^{\lambda',\lambda}\HH_{\zeta,\lambda}\HH^*_{\zeta',\lambda'}
}{
(\sigma_\mu)^{\zeta,\zeta'}}\ .
\end{split}
\end{equation} 
The elements of the real-valued matrix $b_{\rho\nu}$ expressed in terms of amplitudes $\HH$ are:
\begin{equation}
b_{\rho0}=\left(
\begin{array}{c}
| \HH_{--}| ^2+| \HH_{+-}| ^2+| \HH_{-+}| ^2+| \HH_{++}| ^2  \\
2\Re\left({\HH_{++} \HH_{-+}^*}+{\HH_{--} \HH_{+-}^*}\right)  \\
2\Im\left({\HH_{++} \HH_{-+}^*}-{\HH_{--} \HH_{+-}^*}\right) \\
-| \HH_{--}| ^2+| \HH_{++}| ^2+| \HH_{+-}| ^2-| \HH_{-+}| ^2 
\end{array}
\right)  \ ,  
\end{equation}
\begin{equation}
b_{\rho1}=\left(
\begin{array}{c}
2\Re\left({\HH_{++} \HH_{+-}^*}+{\HH_{--} \HH_{-+}^*}
\right)  \\
2\Re\left({\HH_{++} \HH_{--}^*}+
{\HH_{-+} \HH_{+-}^*}\right)  \\
2\Im\left({\HH_{++} \HH_{--}^*}-
{\HH_{-+} \HH_{+-}^*}\right) \\
2\Re\left({\HH_{++} \HH_{+-}^*}-{\HH_{--} \HH_{-+}^*}
\right)
\end{array}
\right)    \ ,  
\end{equation}
\begin{equation}
b_{\rho2}=\left(
\begin{array}{c}
-2\Im\left({\HH_{++} \HH_{+-}^*}-{\HH_{--} \HH_{-+}^*}
\right)  \\
-2\Im\left({\HH_{++} \HH_{--}^*}+{ \HH_{-+}\HH_{+-}^*}\right)  \\
2\Re\left({\HH_{++} \HH_{--}^*}-{ \HH_{-+}\HH_{+-}^*}\right) \\
-2\Im\left({\HH_{++} \HH_{+-}^*}+{\HH_{--} \HH_{-+}^*}\right)
\end{array}
\right)     \ , 
\end{equation}
\begin{equation}
b_{\rho3}=\left(
\begin{array}{c}
-| \HH_{--}| ^2-| \HH_{+-}| ^2+| \HH_{-+}| ^2+| \HH_{++}| ^2  \\
2\Re\left({\HH_{++} \HH_{-+}^*}-
{\HH_{--} \HH_{+-}^*}\right)  \\
2\Im\left({\HH_{++} \HH_{-+}^*}+
{\HH_{--} \HH_{+-}^*}\right) \\
| \HH_{--}| ^2+| \HH_{++}| ^2-| \HH_{+-}| ^2-| \HH_{-+}| ^2 
\end{array}
\right)    \ .
\end{equation}
The matrix elements $b_{\mu\nu}$ are interrelated since they are expressed by the four complex amplitudes $\HH_{\lambda,\lambda'}$. Therefore, neglecting the unobservable overall phase there are up to six independent real-valued functions in addition to the unpolarized cross section term $b_{00}$. The $b$-matrix can be considered as a generalization of Lee-Yang baryon polarization formula~\cite{Lee:1957qs} which has maximum two independent parameters (see example in Appendix~\ref{sec:bb}). The terms $b_{i0}/b_{00}$ are discussed in \cite{LHCb:2023crj} in the context of hadronic decays and are called aligned polarimeter fields $\alpha_{x,y,z}$. In Appendix~\ref{sec:dmexamples} we give  the $b$ matrices for few example processes.

\section{Complete decay matrix for SL decays}\label{app:nf}

The terms of the non-flip contributions for the unaligned (with arbitrary $\phi_l$) decay matrix $b^\textrm{nf}_{\mu\nu}$ are (the term $b_{33}^\textrm{nf}$ does not depend on the angle and it is not repeated): 
\begin{equation}
\begin{aligned}
 b^\textrm{nf}_{11} & = \phantom{-}\Re(\mathcal{E}_{00}^\textrm{nf})+\left\{\Re(\mathcal{E}_{11}^\textrm{nf})\cos2\phi_l-\Im(\mathcal{E}_{11}^\textrm{nf})\sin2\phi_l\right\},\\
 b^\textrm{nf}_{12} & =-\Im(\mathcal{E}_{00}^\textrm{nf})-\left\{\Re(\mathcal{E}_{11}^\textrm{nf})\sin2\phi_l+\Im(\mathcal{E}_{11}^\textrm{nf})\cos2\phi_l\right\},\\
 b^\textrm{nf}_{13} & =\phantom{-}\Re(\mathcal{I}_{13}^{\textrm{nf}})\cos\phi_l-\Im(\mathcal{I}_{13}^{\textrm{nf}})\sin\phi_l, \\
 b^\textrm{nf}_{21} & =\phantom{-}\Im(\mathcal{E}_{00}^\textrm{nf})-\left\{\Re(\mathcal{E}_{11}^\textrm{nf})\sin2\phi_l+\Im(\mathcal{E}_{11}^\textrm{nf})\cos2\phi_l\right\},\\
 b^\textrm{nf}_{22} & =\phantom{-}\Re(\mathcal{E}_{00}^\textrm{nf})-\left\{\Re(\mathcal{E}_{11}^\textrm{nf})\cos2\phi_l-\Im(\mathcal{E}_{11}^\textrm{nf})\sin2\phi_l\right\},\\
 b^\textrm{nf}_{23} & = -(\Re(\mathcal{I}_{13}^{\textrm{nf}})\sin\phi_l+\Im(\mathcal{I}_{13}^{\textrm{nf}})\cos\phi_l), \\
 b^\textrm{nf}_{31} & =-(\Re(\mathcal{I}_{31}^{\textrm{nf}})\cos\phi_l-\Im(\mathcal{I}_{31}^{\textrm{nf}})\sin\phi_l),\\
 b^\textrm{nf}_{32} & =\phantom{-}\Re(\mathcal{I}_{31}^{\textrm{nf}})\sin\phi_l+\Im(\mathcal{I}_{31}^{\textrm{nf}})\cos\phi_l \ .\\
\end{aligned}
\end{equation}
The remaining terms of the flip contributions for the decay matrix  $b^\textrm{f}_{\mu\nu}$ are: 
\begin{equation}
\begin{aligned}
  b_{11}^\textrm{f} & = \phantom{-}\Re(\mathcal{E}_{00}^\textrm{f})-\left\{\Re(\mathcal{E}_{11}^\textrm{f})\cos2\phi_l-\Im(\mathcal{E}_{11}^\textrm{f})\sin2\phi_l\right\},\\
  b_{12}^\textrm{f} & =-\Im(\mathcal{E}_{00}^\textrm{f})+\left\{\Re(\mathcal{E}_{11}^\textrm{f})\sin2\phi_l+\Im(\mathcal{E}_{11}^\textrm{f})\cos2\phi_l\right\},\\
  b_{13}^\textrm{f} & =\phantom{-}\Re(\mathcal{I}_{13}^\textrm{f})\cos\phi_l-\Im(\mathcal{I}_{13}^\textrm{f})\sin\phi_l, \\
  b_{21}^\textrm{f} & =\phantom{-}\Im(\mathcal{E}_{00}^\textrm{f})+\left\{\Re(\mathcal{E}_{11}^\textrm{f})\sin2\phi_l+\Im(\mathcal{E}_{11}^\textrm{f})\cos2\phi_l\right\},\\
  b_{22}^\textrm{f} & =\phantom{-}\Re(\mathcal{E}_{00}^\textrm{f})+\left\{\Re(\mathcal{E}_{11}^\textrm{f})\cos2\phi_l-\Im(\mathcal{E}_{11}^\textrm{f})\sin2\phi_l\right\},\\
 b_{23}^\textrm{f} & = -(\Re(\mathcal{I}_{13}^\textrm{f})\sin\phi_l+\Im(\mathcal{I}_{13}^\textrm{f})\cos\phi_l), \\
 b_{31}^\textrm{f} & =-(\Re(\mathcal{I}_{31}^\textrm{f})\cos\phi_l-\Im(\mathcal{I}_{31}^\textrm{f})\sin\phi_l),\\
 b_{32}^\textrm{f} & = \phantom{-}\Re(\mathcal{I}_{31}^\textrm{f})\sin\phi_l+\Im(\mathcal{I}_{31}^\textrm{f})\cos\phi_l \ .\\
\end{aligned}
\end{equation}

\section{Examples of aligned decay matrices}\label{sec:dmexamples}
\subsection[]{$B_1\to B_2\gamma$}
The amplitude Eq.~\eqref{eq:helHad} for the weak decay $B_1\to B_2\gamma$ simplifies by replacing $\ulambda_W\to\lambda_\gamma$ where $\lambda_\gamma=\{-1,1\}$.
For the hadronic tensor only terms $H_{\frac{1}{2}1}$ and $H_{-\frac{1}{2}-1}$  are non zero. The transition tensor for decay with real photon in helicity representation reads:
\begin{align}
T^{\kappa\kappa',\lambda_2\lambda_2'} &= \frac{1}{4\pi}\sum_{\lambda_\gamma}H_{\lambda_2\lambda_\gamma}H^{\ast}_{\lambda_2'\lambda_\gamma} \mathcal{D}^{1/2\ast}_{\kappa,\lambda_2-\lambda_\gamma}(\Omega_2)\mathcal{D}^{1/2}_{\kappa',\lambda_2'-\lambda_\gamma}(\Omega_2) \ .
\end{align}
The decay matrix $b_{\mu\nu}^{\gamma}$ is following:
\begin{equation}
\begin{split}
b_{\mu\nu}^\gamma:= & \sum_{\lambda_\gamma}\sum_{\lambda_2,\lambda_2'=-1/2}^{1/2}H_{\lambda_2\lambda_\gamma}H^{\ast}_{\lambda_2'\lambda_\gamma}{\sigma_{\mu}^{\lambda_2-\lambda_\gamma,\lambda_2'-\lambda_\gamma}\sigma_{\nu}^{\lambda_2',\lambda_2} }\\
=& 
|H_{-1/2,-1}|^2{\sigma_{\mu}^{1/2,1/2}\sigma_{\nu}^{-1/2,-1/2} }
+ 
|H_{1/2,+1}|^2{\sigma_{\mu}^{-1/2,-1/2}\sigma_{\nu}^{1/2,1/2} }\ 
\end{split}
\end{equation}
or
\begin{equation}
    b_{\mu\nu}^\gamma\propto\left(
\begin{array}{rrrr}
 1 & \phantom{-}0 & \phantom{-}0 & \alpha_\gamma \\
 0 & 0 & 0 & 0 \\
 0 & 0 & 0 & 0 \\
 -\alpha_\gamma & 0 & 0 & -1 \\
\end{array}
\right)
\end{equation}
where
\begin{equation}
    \alpha_\gamma=|H_{1/2,+1}|^2-|H_{-1/2,-1}|^2; \quad {|H_{1/2,+1}|^2+|H_{-1/2,-1}|^2=1} \ .
\end{equation}

\subsection[]{$B_1\to B_2\pi$}\label{sec:bb} For weak non-leptonic decay $D(B_1\to B_2\pi)$ we present the results from Ref.~\cite{Perotti:2018wxm} as a product of rotation matrix and the aligned decay matrix:
\begin{align}
T^{\kappa\kappa',\lambda_2\lambda_2'} &= \frac{1}{4\pi}H_{\lambda_2,0}H^{\ast}_{\lambda_2',0} \mathcal{D}^{1/2\ast}_{\kappa,\lambda_2}(\Omega_2)\mathcal{D}^{1/2}_{\kappa',\lambda_2'}(\Omega_2) \ .
\end{align}
The decay matrix $b_{\mu\nu}^D$ is rewritten as
\begin{equation}
\begin{split}
b_{\mu\nu}^D:= & \sum_{\lambda_2,\lambda_2'=-1/2}^{1/2}H_{\lambda_2,0}H^{\ast}_{\lambda_2',0}{\sigma_{\mu}^{\lambda_2,\lambda_2'}\sigma_{\nu}^{\lambda_2',\lambda_2} }\\
=& |H_{-1/2,0}|^2 \sigma_\mu^{-1/2,-1/2}\sigma_\nu^{-1/2,-1/2} + |H_{1/2,0}|^2 \sigma_\mu^{1/2,1/2}\sigma_\nu^{1/2,1/2} \\
\phantom{=}& + H_{1/2,0}H_{-1/2,0}^* \sigma_\mu^{1/2,-1/2}\sigma_\nu^{-1/2,1/2} + H_{-1/2,0}H_{1/2,0}^* \sigma_\mu^{-1/2,1/2}\sigma_\nu^{1/2,-1/2}
\ 
\end{split}
\end{equation}
or
\begin{equation}
    b_{\mu\nu}^D\propto\left(
\begin{array}{cccc}
 1 & 0 & 0 & \alpha_D \\
 0 & \gamma_D & -\beta_D & 0 \\
 0 & \beta_D & \gamma_D & 0 \\
 \alpha_D & 0 & 0 & 1 \\
\end{array}
\right)
\end{equation}
where
\begin{align}
    \alpha_D=|H_{1/2,0}|^2-|H_{-1/2,0}|^2; &\quad {|H_{1/2,0}|^2+|H_{-1/2,0}|^2=1}, \\
    \beta_D=2\Im(H_{1/2,0}H_{-1/2,0}^*); &\quad \gamma_D = 2\Re(H_{1/2,0}H_{-1/2,0}^*) \ .
\end{align}

\subsection[]{$B_1\to B_2\gamma^*\to B_2 l^+l^-$}\label{sec:cc} The decay matrices for the $B_1\to B_2\gamma^*\to B_2 l^+l^-$ electromagnetic decay can be obtained by simplifying the hadronic tensor by setting to zero all formfactors except for $H^V_{\frac{1}{2}1}=H^V_{-\frac{1}{2}-1}$ and $H^V_{\frac{1}{2}0}=H^V_{-\frac{1}{2}0}$ that are non zero in this parity-conserving process.
The decay $\gamma^*\to l^-l^+$ is described in the $\mathbb{R}_\gamma$ frame where the emission angles of the $l^-$ lepton are $\theta_l$ and $\phi_l$. The value of the lepton momentum in this frame is 
\begin{equation}\label{eq:llmom}
    |{\bf p}_l|=\frac{\sqrt{q^2-4m_l^2}}{2}\ .
\end{equation}

The leptonic tensor for the $\gamma^*$ decay $\lambda_\gamma=\{-1,0,1\}$ with the lepton helicities summed over is: 
\begin{align}
 L_{\lambda_\gamma,\lambda'_\gamma} (q^2,\Omega_l)&:=\sum_{\lambda_+=-1/2}^{1/2} \sum_{\lambda_-=-1/2}^{1/2}\braket{\Omega_-,\lambda_-,\lambda_{+}|S_l|,q^2,\lambda'_\gamma}^*\braket{\Omega_-,\lambda_-,\lambda_{+}|S_l|q^2,\lambda_\gamma}\\
 &=\sum_{\lambda_+=-1/2}^{1/2} \sum_{\lambda_-=-1/2}^{1/2}|h^l_{\lambda_+\lambda_{-}}(q^2)|^2\mathcal{D}^{1\ast}_{\lambda_\gamma,\lambda_--\lambda_{+}}(\Omega_l)\mathcal{D}^{1}_{\lambda_\gamma',\lambda_--\lambda_{+} }(\Omega_l)\\
 &=e^{i(\lambda_\gamma-\lambda'_\gamma)\phi_l}\sum_{\lambda_+=-1/2}^{1/2} \sum_{\lambda_-=-1/2}^{1/2}|h^l_{\lambda_+\lambda_{-}}(q^2)|^2d^{1}_{\lambda_\gamma,\lambda_--\lambda_{+}}(\theta_l)d^{1}_{\lambda_\gamma',\lambda_--\lambda_{+}}(\theta_l)\ . 
\end{align}
The moduli squared of $h^l_{\lambda_-\lambda_{+}}$ corresponding to the vertex $\bar{u}(p_z,\lambda_-)\gamma^\mu v(-p_z,\lambda_+)\epsilon_\mu$ calculated using the charged-lepton spinor representation from Appendix in Ref.~\cite{Perdrisat:2006hj} are:
\begin{align} 
    \mathrm{nonflip} (\lambda_\gamma=\mp1)& :\  |h^l_{\lambda_-=\mp\frac{1}{2},\lambda_{+}=\pm\frac{1}{2}}|^2={2q^2},\\
    \text{flip} (\lambda_\gamma=0)& :\  |h^l_{\lambda_-=\pm\frac{1}{2},\lambda_{+}=\pm\frac{1}{2}}|^2={4m_l^2}.
\end{align}
The resulting leptonic tensor reads
\begin{equation}
    \begin{split}
& L_{\lambda_\gamma,\lambda'_\gamma} (q^2,\Omega_l)=\\
 &=   \left( q^2-4 m_l^2\right)\left(
\begin{array}{ccc}
 \cos^2\theta_l & -\sqrt{2} e^{-i \phi_l} \sin \theta_l \cos \theta_l & e^{-2 i \phi_l} \sin ^2\theta_l \\
 {-}\sqrt{2} e^{i  \phi_l} \sin\theta_l\cos \theta_l & -\cos 2 \theta_l & \sqrt{2} e^{-i  \phi_l} \sin \theta_l \cos \theta_l \\
 e^{2 i  \phi_l} \sin ^2\theta_l & \sqrt{2} e^{i  \phi_l} \sin\theta_l \cos \theta_l & \cos^2\theta_l\\
\end{array}
\right)\\
&+\left( q^2+4 m_l^2\right)\text{diag}(1,1,1)
\ .\end{split}
\end{equation}

The differential decay rate of the unpolarized baryon $B_1$ in the electromagnetic conversion process where the spins of all final particles are summed is
\begin{align}
    {\dd\Gamma}
    &\propto \frac{\alpha_{\text{em}}^2}{q^2} V_{Ph}(q^2)\left(1-\frac{4 m_l^2}{q^2}\right)b^{\textrm{em}}_{00}{\dd q \dd\Omega_2 \dd\Omega_l}\ ,
\end{align}
where $V_{Ph}(q^2)$ is the three-body phase space density factor given by the product of the momenta $|{\bf p}_2|$ and $|{\bf p}_l|$ of the baryon $B_2$ and the lepton, given in Eqs.~\eqref{eq:momentum_b2} and~\eqref{eq:llmom}, respectively. The unrotated decay matrix can be obtained adapting~\eqref{eq:bmunu_semil}:
\begin{equation}
\begin{split}
\bem_{\mu\nu}:= & \frac{1}{{2(q^2-4m_l^2)}} \sum_{\lambda_\gamma,\lambda_\gamma'=-1,0}^1\sum_{\lambda_2,\lambda_2'=-1/2}^{1/2}H_{\lambda_2\lambda_\gamma}H^{\ast}_{\lambda_2'\lambda_\gamma'}{\cal T}^{\lambda_\gamma,\lambda_\gamma',\lambda_2,\lambda_2'}_{\mu\nu} \ .
\end{split}
\end{equation}
Its elements are 
\begin{equation}
    \bem_{\mu\nu}=\left(
\begin{array}{cccc}
 \bem_{00} & \bem_{01} & \bem_{02} & 0 \\
 \bem_{01} & \bem_{11}& \bem_{12} & \bem_{13} \\
 \bem_{02} & \bem_{12} & \bem_{22} & \bem_{23} \\
 0 & -\bem_{13} & -\bem_{23} & \bem_{33} \\
\end{array}
\right) \ ,   
\end{equation}
where
\begin{equation}
\begin{split}
  b^{\textrm{em}}_{00}&= \phantom{-}\left[\cos^2\theta_l +\frac{q^2+4 m_l^2 }{q^2-4 m_l^2}\right]|H^V_{\frac{1}{2}1}|^2+2\left[\sin^2\theta_l +\frac{4 m_l^2 }{q^2-4 m_l^2}\right]|H^V_{\frac{1}{2}0}|^2,\\
  b^{\textrm{em}}_{33}&= -\left[\cos^2\theta_l +\frac{q^2+4 m_l^2 }{q^2-4 m_l^2}\right]|H^V_{\frac{1}{2}1}|^2+2\left[\sin^2\theta_l +\frac{4 m_l^2 }{q^2-4 m_l^2}\right]|H^V_{\frac{1}{2}0}|^2,\\
  b^{\textrm{em}}_{01}&=-\sqrt{2} \sin 2\theta_l  \sin\phi_l\Im(H^V_{\frac{1}{2}1}H^{V\ast}_{\frac{1}{2}0}),\\
  b^{\textrm{em}}_{02}&=-\sqrt{2} \sin 2\theta_l  \cos\phi_l\Im(H^V_{\frac{1}{2}1}H^{V\ast}_{\frac{1}{2}0}),\\ 
  \bem_{11}&=\phantom{-}\cos 2\phi_l\sin^2\theta_l |H^V_{\frac{1}{2}1}|^2+2\left[\sin^2\theta_l +\frac{4 m_l^2 }{q^2-4 m_l^2}\right]|H^V_{\frac{1}{2}0}|^2,\\
  \bem_{22}&={-}\cos 2\phi_l\sin^2\theta_l |H^V_{\frac{1}{2}1}|^2+2\left[\sin^2\theta_l +\frac{4 m_l^2 }{q^2-4 m_l^2}\right]|H^V_{\frac{1}{2}0}|^2,\\
  \bem_{12}&=-\sin 2\phi_l\sin^2\theta_l |H^V_{\frac{1}{2}1}|^2,\\
  \bem_{23}&=-\sqrt{2} \sin 2\theta_l  \sin\phi_l\Re(H^V_{\frac{1}{2}1}H^{V\ast}_{\frac{1}{2}0}),\\
  \bem_{13}&=\sqrt{2} \sin 2\theta_l  \cos\phi_l\Re(H^V_{\frac{1}{2}1}H^{V\ast}_{\frac{1}{2}0})  \ .
  \end{split}
\end{equation}
Decay plane aligned parameters reduce to the following form
\begin{equation}
    \bem_{\mu\nu}=\left(
\begin{array}{cccc}
 \bem_{00} & 0 & \bem_{02} & 0 \\
 0& \bem_{11}& 0& \bem_{13} \\
 \bem_{02} & 0& \bem_{22} & 0 \\
 0 & -\bem_{13} & 0& \bem_{33} \\
\end{array}
\right) \ ,   
\end{equation}
where in the real formfactors limit additionally the term $\bem_{02}$ vanishes. Thus, no polarization is induced, but the initial polarization and spin correlations of the baryon $B_1$ are transferred to the daughter baryon.


\subsection[]{$B_1\to B_2 [V^* \to P_1P_2]$} Here we consider a decay of spin-1/2 baryon to a spin-1/2 baryon and a pair of pseudoscalar mesons $P_1$ and $P_2$  via an 
intermediate vector meson $V$ e.g. $B_1\to B_2 \rho^0 \to \pi^+\pi^-$.  The decay matrices are obtained as in Appendix~\ref{sec:cc} by replacing the dilepton with the pseudoscalars, and the virtual photon with a massive vector meson decaying strongly. Since the initial baryon decays weakly into the intermediate state $B_1\to B_2 V^*$, all vector and axial vector formfactors should be used. The decay $V^*(q)\to P_1(m_1,{\bf p}_\pi)P_2(m_2,-{\bf p}_\pi)$ is described in the $\mathbb{R}_V$ frame where the emission angles of the $P_1$ pseudoscalar are $\theta_\pi$ and $\phi_\pi$. The value of the momentum ${\bf p}_\pi$  is
\begin{equation}
    |{\bf p}_\pi|=\sqrt{\frac{q^4+m_{2}^4+m_{1}^4-2q^2 m_1^2-2 q^2 m_2^2-2 m_1^2m_2^2}{4q^2}}\ .
\end{equation}
The tensor for the $V^*\to P_1P_2$ decay for the helicities $\lambda_V,\lambda_V'=\{-1,0,1\}$  is: 
\begin{align}
 \mathfrak{H}_{\lambda_V\lambda'_V} (\Omega_\pi)&:= e^{i(\lambda_V-\lambda'_V)\phi_\pi}|h^V|^2 d^{1}_{\lambda_V,0}(\theta_\pi)d^{1}_{\lambda_V',0}(\theta_\pi)\ , 
\end{align}
where the  $h^V$ is a constant and  it can be absorbed as a normalization factor. The resulting tensor reads
\begin{equation}
    \begin{split}
& \mathfrak{H}_{\lambda_V\lambda'_V} (\Omega_\pi)=\left(
\begin{array}{ccc}
 \frac{\sin^2\theta_\pi}{2} & \frac{e^{-i \phi_\pi} \sin \theta_\pi \cos \theta_\pi}{\sqrt{2}} & -\frac{e^{-2 i \phi_\pi}}{2}\sin ^2\theta_\pi \\
 \frac{e^{i \phi_\pi} \sin \theta_\pi \cos \theta_\pi}{\sqrt{2}} & \cos^2 \theta_\pi & -\frac{e^{-i \phi_\pi} \sin \theta_\pi \cos \theta_\pi}{\sqrt{2}} \\
 -\frac{e^{2 i  \phi_\pi}}{2} \sin ^2\theta_\pi & -\frac{e^{i \phi_\pi} \sin \theta_\pi \cos \theta_\pi}{\sqrt{2}} & \frac{\sin^2\theta_\pi}{2} \\
\end{array}
\right)
\ .\end{split}
\end{equation}
The unrotated decay matrix can be obtained by replacing the leptonic tensor with the tensor $\mathfrak{H}_{\lambda_V,\lambda'_V} $ in~\eqref{eq:bmunu_semil}:
\begin{equation}
\begin{split}
b^V_{\mu\nu}:= & \sum_{\lambda_V,\lambda_V'=-1}^1 H_{\lambda_2\lambda_V}H^{\ast}_{\lambda_2'\lambda_V'}{\cal T}^{\lambda_V,\lambda_V',\lambda_2,\lambda_2'}_{\mu\nu} \ .
\end{split}
\end{equation}
Its elements are 
\begin{equation}
    \begin{split}
    b^V_{00} &= \left(|H_{\frac{1}{2}0}|^2+|H_{-\frac{1}{2}0}|^2\right) \cos^2\theta_\pi+\frac{1}{2}\left(|H_{\frac{1}{2}1}|^2+|H_{-\frac{1}{2}-1}|^2\right)\sin^2\theta_\pi, \\
    b^V_{01} &=  \Re(\mathcal{A})\cos\phi_\pi +  \Im(\mathcal{A})\sin\phi_\pi, \\
    b^V_{02} &=  \Im(\mathcal{A})\cos\phi_\pi -  \Re(\mathcal{A})\sin\phi_\pi, \\
    b^V_{03} &= \left(|H_{\frac{1}{2}0}|^2-|H_{-\frac{1}{2}0}|^2\right) \cos^2\theta_\pi+\frac{1}{2}\left(|H_{\frac{1}{2}1}|^2-|H_{-\frac{1}{2}-1}|^2\right)\sin^2\theta_\pi, \\
    b^V_{10} &=  \Re(\mathcal{B})\cos\phi_\pi +  \Im(\mathcal{B})\sin\phi_\pi, \\
    b^V_{20} &=  \Im(\mathcal{B})\cos\phi_\pi -  \Re(\mathcal{B})\sin\phi_\pi, \\
    b^V_{11} &= \Re(\mathcal{C}) -\Re(\mathcal{D})\cos{2\phi_\pi} -  \Im(\mathcal{D})\sin{2\phi_\pi}, \\
    b^V_{12} &= \Im(\mathcal{C}) - \Im(\mathcal{D})\cos{2\phi_\pi} +  \Re(\mathcal{D})\sin{2\phi_\pi}, \\
    b^V_{21} &= -\Im(\mathcal{C}) -\Im(\mathcal{D})\cos{2\phi_\pi}  + \Re(\mathcal{D}) \sin{2\phi_\pi}, \\
    b^V_{22} &= \Re(\mathcal{C}) + \Re(\mathcal{D})\cos{2\phi_\pi}  + \Im(\mathcal{D}) \sin{2\phi_\pi}, \\
    b^V_{13} &= -\Re(\mathcal{E})\cos\phi_\pi  -  \Im(\mathcal{E})\sin\phi_\pi, \\
    b^V_{23} &= -\Im(\mathcal{E})\cos\phi_\pi  + \Re(\mathcal{E}) \sin\phi_\pi, \\
    b^V_{30} &= \left(|H_{\frac{1}{2}0}|^2-|H_{-\frac{1}{2}0}|^2\right) \cos^2\theta_\pi -\frac{1}{2}\left(|H_{\frac{1}{2}1}|^2-|H_{-\frac{1}{2}-1}|^2\right) \sin^2\theta_\pi, \\
    b^V_{31} &= \Re(\mathcal{F})\cos\phi_\pi  + \Im(\mathcal{F}) \sin\phi_\pi, \\
    b^V_{32} &= \Im(\mathcal{F})\cos\phi_\pi  - \Re(\mathcal{F}) \sin\phi_\pi, \\
    b^V_{33} &= \left(|H_{\frac{1}{2}0}|^2+|H_{-\frac{1}{2}0}|^2\right) \cos^2\theta_\pi -\frac{1}{2}\left(|H_{\frac{1}{2}1}|^2+|H_{-\frac{1}{2}-1}|^2\right)\sin^2\theta_\pi,
\end{split}
\end{equation}
with
\begin{align}
    \mathcal{A} &= \sqrt{2} \cos\theta_\pi \sin\theta_\pi \left(H_{\frac{1}{2}0}^* H_{-\frac{1}{2}-1}-H_{\frac{1}{2}1}^* H_{-\frac{1}{2}0}\right), \\
    \mathcal{B} &= \sqrt{2} \cos\theta_\pi \sin\theta_\pi \left(H_{-\frac{1}{2}0}^* H_{-\frac{1}{2}-1}-H_{\frac{1}{2}1}^* H_{\frac{1}{2}0}\right), \\
    \mathcal{C} &= 2 H_{\frac{1}{2}0}^* H_{-\frac{1}{2}0} \cos^2\theta_\pi, \\
    \mathcal{D} &= H_{\frac{1}{2}1}^* H_{-\frac{1}{2}-1} \sin^2\theta_\pi, \\
    \mathcal{E} &= \sqrt{2} \cos\theta_\pi \sin\theta_\pi \left(H_{-\frac{1}{2}0}^* H_{-\frac{1}{2}-1}+H_{\frac{1}{2}1}^* H_{\frac{1}{2}0}\right), \\
    \mathcal{F} &= \sqrt{2} \cos\theta_\pi \sin\theta_\pi \left(H_{\frac{1}{2}0}^* H_{-\frac{1}{2}-1}+H_{\frac{1}{2}1}^* H_{-\frac{1}{2}0}\right).
\end{align}
Decay plane aligned parameters reduce to the following form
\begin{equation}
    b^V_{\mu\nu}=\left(
\begin{array}{cccc}
 b^V_{00} & \Re(\mathcal{A}) & \Im(\mathcal{A}) & b^V_{03} \\
 \Re(\mathcal{B}) & \Re(\mathcal{C-D})& \Im(\mathcal{C-D}) & -\Re(\mathcal{E}) \\
 \Im(\mathcal{B}) & -\Im(\mathcal{C+D}) & \Re(\mathcal{C+D}) & -\Im(\mathcal{E}) \\
 b^V_{30} & \Re(\mathcal{F}) & \Im(\mathcal{F}) & b^V_{33} \\
\end{array}
\right) \ .
\end{equation}
The differential decay rate of the process with unpolarized baryon $B_1$ and the spins of $B_2$ summed over is
\begin{align}
    {\dd\Gamma}
    &\propto  V_{Ph}(q^2)b^V_{00}{\dd q \dd\Omega_2 \dd\Omega_\pi}\ ,
\end{align}
where $V_{Ph}(q^2)$ is the three-body phase space density factor given by the product of the momenta $|{\bf p}_2|$ and $|{\bf p}_\pi|$.
\bibliographystyle{apsrev4-1} 
\bibliography{refBB}

\end{document}